\documentclass[11pt,nofootinbib,floatfix,onecolumn,preprintnumbers]{revtex4}

\usepackage{amsmath,ulem}
\usepackage{amssymb}
\usepackage{graphicx}
\usepackage{rotating}
\usepackage{color}
\usepackage{multirow} 
\usepackage{fontenc} 
\usepackage{slashed}
\usepackage[mathscr]{eucal}
\usepackage{color}
\usepackage{longtable}
\usepackage{ulem}
\usepackage{pifont} 
\usepackage{hhline}

\newcommand{\AddrTakashi}{%
Institute for Particle Physics Phenomenology, 
University of Durham, Durham, DH1 3LE, United Kingdom
}

\newcommand{\AddrAvelino}{%
IFPA, Dep. AGO, Universit\'e de Li\`ege, Bat B5, Sart-Tilman B-4000
Li\`ege 1, Belgium 
}

\preprint{IPPP/13/99}
\preprint{DCPT/13/198}
\begin{document}

\title{Lepton Flavor Violation in the Scotogenic Model}

\author{Takashi Toma}\email{takashi.toma@durham.ac.uk}
\affiliation{\AddrTakashi}

\author{Avelino Vicente} \email{avelino.vicente@ulg.ac.be}
\affiliation{\AddrAvelino}

\begin{abstract}
We investigate lepton flavor violation in the scotogenic model
proposed by Ma in which neutrinos acquire non-zero masses at the
1-loop level. Although some works exist in this direction, they have
mainly focused on the radiative decay $\ell_\alpha \to \ell_\beta
\gamma$. Motivated by the promising new projects involving other
low-energy processes, we derive complete analytical expressions for
$\ell_\alpha \to 3 \, \ell_\beta$ and $\mu-e$ conversion in nuclei,
and numerically study their impact on the phenomenology. We will show
that these processes can actually have rates larger than the one for
$\ell_\alpha \to \ell_\beta \gamma$, thus providing more stringent
constraints and better experimental perspectives.
\end{abstract}

\maketitle

\section{Introduction}
\label{sec:intro}

The search for lepton flavor violation (LFV) is going to live an
unprecedented era with great experimental efforts in many different
fronts. In addition to the well-known searches for the radiative decay
$\ell_\alpha \to \ell_\beta \gamma$, new projects involving other low-energy
processes, such as $\ell_\alpha \to 3 \, \ell_\beta$ or $\mu-e$ conversion in
nuclei, are going to look for a positive LFV signal. 

For many years, the experiment leading to the most stringent constraints
has been MEG \cite{Adam:2011ch}. This experiment, which searches for
the radiative decay $\mu \to e \gamma$, recently published a new
limit, $\text{Br}(\mu \to e \gamma) < 5.7 \times 10^{-13}$, obtained
with an updated analysis of the 2009-2010 data sample together with
the analysis of the new data collected in 2011
\cite{Adam:2013mnn}. The expectation is that MEG can reduce the
current bound by another order of magnitude, with sensitivities of
about $6 \times 10^{-14}$ after 3 years of acquisition time
\cite{Baldini:2013ke}.

However, the most impressive improvements in the next few years are
expected in $\mu \to 3 e$ and $\mu-e$ conversion in
nuclei. For the former, the Mu3e experiment is expected to reach a
sensitivity of $10^{-15}$ (after upgrades $10^{-16}$)
\cite{Blondel:2013ia}. This would imply an improvement of 3-4 orders
of magnitude with respect to the current bound. For $\mu-e$ conversion
in nuclei several project will compete in the next few years. These
include Mu2e \cite{Glenzinski:2010zz,Carey:2008zz}, DeeMe
\cite{Aoki:2010zz}, COMET \cite{Cui:2009zz} and PRISM/PRIME
\cite{PRIME}. The expected sensitivities for the conversion rate range
from a modest $10^{-14}$ to an impressive $10^{-18}$.

Finally, the limits for $\tau$ observables are less stringent, although
significant improvements are expected at $B$ factories
\cite{O'Leary:2010af,Hayasaka:2013dsa}. Table \ref{tab:sensi} summarizes the
current experimental bounds and future sensitivities for the
low-energy LFV observables.

\begin{table}[tb!]
\centering
\begin{tabular}{|c|c|c|}
\hline
LFV Process & Present Bound & Future Sensitivity  \\
\hhline{|=|=|=|}
$\mu \to e \gamma$ & $5.7 \times 10^{-13}$ \cite{Adam:2013mnn} & $6 \times 10^{-14}$ \cite{Baldini:2013ke}  \\
$\tau \to e \gamma$ & $3.3 \times 10^{-8}$ \cite{Aubert:2009ag}& $\sim 10^{-8}-10^{-9}$ \cite{Hayasaka:2013dsa}\\
$\tau \to \mu \gamma$ & $4.4 \times 10^{-8}$ \cite{Aubert:2009ag}& $\sim 10^{-8}-10^{-9}$ \cite{Hayasaka:2013dsa} \\
$\mu \to 3 e$ & $1.0 \times 10^{-12}$\cite{Bellgardt:1987du} & $\sim 10^{-16}$ \cite{Blondel:2013ia}\\
$\tau \to 3 e$ & $2.7\times10^{-8}$\cite{Hayasaka:2010np} & $\sim 10^{-9}-10^{-10}$ \cite{Hayasaka:2013dsa}  \\
$\tau \to 3 \mu$ & $2.1\times10^{-8}$\cite{Hayasaka:2010np} & $\sim 10^{-9}-10^{-10}$ \cite{Hayasaka:2013dsa}  \\
$\mu^-$, Au $\to$ $e^-$, Au & $7.0 \times 10^{-13}$ \cite{Bertl:2006up} & $-\!\!\!-\!\!\!-$  \\
$\mu^-$, Ti $\to$ $e^-$, Ti & $4.3 \times 10^{-12}$ \cite{Dohmen:1993mp} & $\sim 10^{-18}$ \cite{PRIME} \\
\hline
\end{tabular}
\caption{Current experimental bounds and future sensitivities for some
  low-energy LFV observables.}
\label{tab:sensi}
\end{table}

Different observables may have very different rates for a given
model. For example, the rates for $\mu \to 3 e$ and $\mu-e$ conversion
in nuclei are typically suppressed with respect to $\mu \to e \gamma$
in models where the dominant LFV contributions are induced by dipole
operators, like the Minimal Supersymmetric Standard Model. However,
there are many frameworks where this is not the case. For this reason,
one needs to fully understand the \textit{anatomy} of LFV in each model
in order to determine the expected hierarchies among observables,
which then become indirect tests of the model.

In this paper we pursue this goal in the context of a model proposed
by Ma in which neutrinos acquire masses at the 1-loop
level~\cite{Ma:2006km}. The same symmetry that forbids the tree-level
contribution to Dirac neutrino masses, a $\mathbb{Z}_2$ parity, also
gives rise to a dark matter candidate. This simple extension of the
Standard Model (SM), usually called \textit{Scotogenic Model},
constitutes a very simple framework to address the most important
motivations to go beyond\footnote{For other recent works on further
  extended models with radiative neutrino masses, see for
  example~\cite{Okada:2012np, Dev:2012sg, Aoki:2013gzs,
    Kajiyama:2013rla, Kanemura:2013qva, Law:2013saa, Hirsch:2013ola,
    Restrepo:2013aga, Ma:2013yga, Lindner:2013awa, Okada:2013iba}.}.
Although some works have been already done regarding LFV in this
model~\cite{Kubo:2006yx, Sierra:2008wj, Suematsu:2009ww,
  Adulpravitchai:2009gi}, they have either focused on $\mu \to e
\gamma$ or neglected contributions beyond the photonic dipole. To the
best of our knowledge, this is the first time $\ell_\alpha \to 3 \,
\ell_\beta$ and $\mu-e$ conversion in nuclei are fully considered. As
we will see, these processes might actually have rates larger than the
one for $\mu \to e \gamma$, thus providing better bounds and
experimental perspectives.

The rest of the paper is organized as follows: in Sec. \ref{sec:model}
we describe the model and its basic features. In
Sec. \ref{sec:analytical} we present our analytical results, whereas
Sec. \ref{sec:pheno} contains a numerical discussion addressing some
phenomenological issues of interest. Finally, we summarize our results
and conclude in Sec. \ref{sec:conclusions}.

\section{The model}
\label{sec:model}

The model under consideration \cite{Ma:2006km} adds three right-handed
neutrinos $N_i$ ($i=1$-$3$) and one $SU(2)_L$ doublet $\eta$ to the SM
particle content. In addition, a $\mathbb{Z}_2$ parity is imposed,
under which the new particles are odd and the SM ones are
even\footnote{Due to the conservation of the $\mathbb{Z}_2$ symmetry,
  the left-handed neutrinos in the SM lepton doublet do not form a
  Dirac pair with the `right-handed' neutrinos $N_i$. For this reason,
  strictly speaking, it is not correct to call the $N_i$ singlets
  right-handed neutrinos. Nevertheless, this has become common
  practice in the literature and we will stick to this
  denomination. \label{foot:N}}. The interaction of the right-handed
neutrino sector is described by the Lagrangian
\begin{equation}
\mathcal{L}_N=\overline{N_i}\partial\!\!\!/N_i
-\frac{m_{N_i}}{2}\overline{N_i^c}P_RN_i+
y_{i\alpha}\eta\overline{N_i}P_L\ell_\alpha+\mathrm{h.c.}. 
\end{equation}
Note that one can always write the right-handed neutrino mass term as
a diagonal matrix without loss of generality.  The scalar potential
$\mathcal{V}$ is given by
\begin{eqnarray}
\mathcal{V}\!\!\!&=&\:
m_{\phi}^2\phi^\dag\phi+m_\eta^2\eta^\dag\eta+
\frac{\lambda_1}{2}\left(\phi^\dag\phi\right)^2+
\frac{\lambda_2}{2}\left(\eta^\dag\eta\right)^2+
\lambda_3\left(\phi^\dag\phi\right)\left(\eta^\dag\eta\right)\nonumber\\
&&\!\!\!+
\lambda_4\left(\phi^\dag\eta\right)\left(\eta^\dag\phi\right)+
\frac{\lambda_5}{2}\left[\left(\phi^\dag\eta\right)^2+
\left(\eta^\dag\phi\right)^2\right] \, .
\end{eqnarray}
We assume that the parameters in the scalar potential are such that
the doublet $\eta$ does not get a vacuum expectation value. This is
fundamental in order to keep the $\mathbb{Z}_2$ symmetry unbroken.
After electroweak symmetry breaking, the masses of the charged component
$\eta^+$ and neutral component
$\eta^0=(\eta_R+i\eta_I)/\sqrt{2}$ are split to
\begin{eqnarray}
m_{\eta^+}^2&=&m_\eta^2+\lambda_3\langle\phi^0\rangle^2\\
m_R^2&=&m_{\eta}^2+\left(\lambda_3+\lambda_4+\lambda_5\right)\langle\phi^0\rangle^2,\\
m_I^2&=&m_{\eta}^2+\left(\lambda_3+\lambda_4-\lambda_5\right)\langle\phi^0\rangle^2,
\end{eqnarray}
where the mass difference between $\eta_R$ and $\eta_I$ is
$m_R^2-m_I^2=2\lambda_5\langle\phi^0\rangle^2$. 

After symmetry breaking, the light neutrino masses are generated at the 1-loop level\footnote{Note that the tree-level contribution is actually forbidden by the $\mathbb{Z}_2$ discrete symmetry.}. The neutrino mass matrix can be expressed as
\begin{eqnarray}
\left(m_{\nu}\right)_{\alpha\beta}&=&
\sum_{i=1}^3\frac{y_{i\alpha}y_{i\beta}}{(4\pi)^2}m_{N_i}
\left[\frac{m_R^2}{m_R^2-m_{N_i}^2}\log\left(\frac{m_R^2}{m_{N_i}^2}\right)
-\frac{m_I^2}{m_I^2-m_{N_i}^2}\log\left(\frac{m_I^2}{m_{N_i}^2}\right)\right]\nonumber\\
&\equiv&\left(y^{T}\Lambda y\right)_{\alpha\beta},
\label{eq:nu-mass}
\end{eqnarray}
where $m_R$ and $m_I$ are the masses of $\eta_R$ and $\eta_I$
respectively, and the $\Lambda$ matrix is defined as 
\begin{equation}
\Lambda=\left(
\begin{array}{ccc}
\Lambda_1 & 0         & 0\\
 0        & \Lambda_2 & 0\\
 0        & 0         & \Lambda_3
\end{array}
\right),\quad
\Lambda_i=\frac{m_{N_i}}{(4\pi)^2}
\left[\frac{m_R^2}{m_R^2-m_{N_i}^2}\log\left(\frac{m_R^2}{m_{N_i}^2}\right)
-\frac{m_I^2}{m_I^2-m_{N_i}^2}\log\left(\frac{m_I^2}{m_{N_i}^2}\right)\right] \, .
\end{equation}
In particular, when $m_R^2 \approx m_I^2 \equiv m_0^2$ ($\lambda_5\ll1$),
the mass matrix gets the simplified form
\begin{equation}
\left(m_\nu\right)_{\alpha\beta}\approx
\sum_{i=1}^3\frac{2\lambda_5 y_{i\alpha}y_{i\beta}\langle\phi^0\rangle^2}
{(4\pi)^2m_{N_i}}
\left[\frac{m_{N_i}^2}{m_{0}^2-m_{N_i}^2}
+\frac{m_{N_i}^4}{\left(m_{0}^2-m_{N_i}^2\right)^2}
\log\left(\frac{m_{N_i}^2}{m_0^2}\right)\right].
\end{equation}
This neutrino mass matrix is diagonalized as
\begin{equation}
U_{\mathrm{PMNS}}^{T} \, m_{\nu} \, U_{\mathrm{PMNS}}=\hat{m}_{\nu}\equiv
\left(
\begin{array}{ccc}
m_1 & 0 & 0\\
0 & m_2 & 0\\
0 & 0 & m_3
\end{array}
\right) \, ,
\end{equation}
where 
\begin{equation}
\label{eq:PMNS}
U_{\mathrm{PMNS}}=
\left(
\begin{array}{ccc}
 c_{12}c_{13} & s_{12}c_{13}  & s_{13}e^{i\delta}  \\
-s_{12}c_{23}-c_{12}s_{23}s_{13}e^{-i\delta}  & 
c_{12}c_{23}-s_{12}s_{23}s_{13}e^{-i\delta}  & s_{23}c_{13}  \\
s_{12}s_{23}-c_{12}c_{23}s_{13}e^{-i\delta}  & 
-c_{12}s_{23}-s_{12}c_{23}s_{13}e^{-i\delta}  & c_{23}c_{13}  
\end{array}
\right) \times
\left(
\begin{array}{ccc}
e^{i\varphi_1/2} & 0 & 0 \\
0 & e^{i\varphi_2/2}  & 0 \\
0 & 0 & 1
\end{array}
\right)
\end{equation}
is the PMNS (Pontecorvo-Maki-Nakagawa-Sakata) matrix. Here $c_{ij} =
\cos \theta_{ij}$, $s_{ij} = \sin 
\theta_{ij}$, $\delta$ is the Dirac phase and $\varphi_1$, $\varphi_2$ are
the Majorana phases\footnote{We will neglect Majorana phases in all
  our computations.}.

The Yukawa matrix $y_{i\alpha}$ can be written using an adapted Casas-Ibarra
parametrization~\cite{Casas:2001sr} as
\begin{equation}
\label{eq:casas-ibarra}
y=\sqrt{\Lambda}^{-1}R\sqrt{\hat{m}_\nu}U_{\mathrm{PMNS}}^{\dag}.
\end{equation}
where $R$ is an complex orthogonal matrix which satisfies $R^T R=1$.

\section{Analytical results}
\label{sec:analytical}

In this section we present our analytical results for the LFV
processes $\ell_\alpha\to\ell_\beta\gamma$,
$\ell_\alpha \to 3 \, \ell_\beta$ and $\mu-e$
conversion in nuclei.  Before we proceed to the analytical discussion
a comment is in order. It is well-known that the rates for LFV
processes get greatly enhanced in models with right-handed neutrinos
at the electroweak scale
\cite{Ilakovac:1994kj,Deppisch:2004fa,Deppisch:2005zm,Ilakovac:2009jf,Alonso:2012ji,Dinh:2012bp,Ilakovac:2012sh,Abada:2012cq,Dev:2013oxa}. This
is due to the fact that the GIM suppression at work in the SM
contribution is spoiled by the mixing between left- and right-handed
neutrinos. One could naively think that this is also the case in the
scotogenic model. However, the unbroken $\mathbb{Z}_2$ symmetry
forbids this mixing, (see footnote \ref{foot:N}), and thus the
enhancement in the $W-\nu$ loops is not present. We will show that the
enhancement is still possible, but with $\eta^\pm-N$ loops instead.

\subsection{$\ell_\alpha\to\ell_\beta\gamma$}

\begin{figure}[t]
\begin{center}
\includegraphics[scale=0.7]{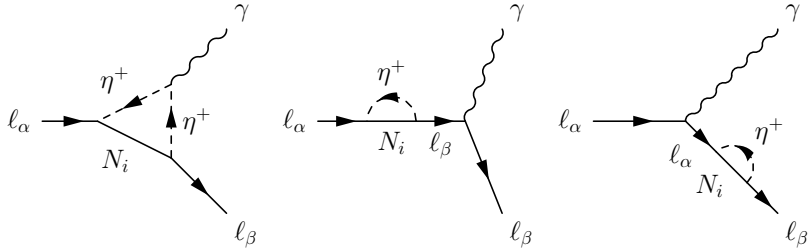}
\caption{1-loop Feynman diagrams leading to
  $\ell_\alpha\to\ell_\beta\gamma$.}
\label{fig:meg}
\end{center}
\end{figure}

The most popular searches for LFV have focused on the radiative
process $\ell_\alpha\to\ell_\beta\gamma$. This is described by the
effective Lagrangian
\begin{equation}
\mathcal{L}_{\mathrm{eff}}=
\left(\frac{\mu_{\beta\alpha}}{2}\right)\overline{\ell_\beta}\sigma^{\mu\nu}
\ell_{\alpha}F_{\mu\nu},
\end{equation}
where $\mu_{\beta\alpha}$ is a transition magnetic moment. It proves
convenient to define it in terms of the dipole form factor $A_D$ as
$\mu_{\beta\alpha}=e m_\alpha A_D/2$, where terms proportional to
$m_{\beta}$ have been neglected and $e$ is the electromagnetic
coupling, related to the electromagnetic fine structure constant as
$\alpha_{\mathrm{em}}=e^2/(4\pi)$. In the model under consideration,
$A_D$ gets contributions at the 1-loop level from the Feynman diagrams
in Fig.~\ref{fig:meg}. They lead to the following expression
\begin{equation}
A_D = \sum_{i=1}^3\frac{y_{i\beta}^*y_{i\alpha}}
{2(4\pi)^2}\frac{1}{m_{\eta^+}^2}
F_2\left(\xi_i\right) \, ,
\label{eq:A2R}
\end{equation}
where the $\xi_i$ parameters are defined as $\xi_i\equiv
m_{N_i}^2/m_{\eta^+}^2$ and the loop function $F_2(x)$ is given in
appendix \ref{sec:appendix1}.
Finally, the branching fraction for $\ell_\alpha \to \ell_\beta
\gamma$ is calculated as
\begin{equation}
\mathrm{Br}\left(\ell_{\alpha}\to\ell_{\beta}\gamma\right)=
\frac{3(4\pi)^3 \alpha_{\mathrm{em}}}{4G_F^2} 
|A_D|^2
\mathrm{Br}\left(\ell_{\alpha}\to\ell_{\beta}\nu_{\alpha}
\overline{\nu_{\beta}}\right) \, ,
\end{equation}
where $G_F$ is the Fermi constant.

\subsection{$\ell_\alpha \to 3 \, \ell_\beta$}
\label{subsec:l3l}

Next we consider the process $\ell_\alpha \to 3 \, \ell_\beta$ (more
precisely denoted as
$\ell_\alpha\to\ell_\beta\bar{\ell}_\beta\ell_\beta$). Although this
has attracted less attention, important projects are going to be
launched in the near future, with the Mu3e experiment as the leading
one.
There are four types of 1-loop diagrams that contribute to
$\ell_\alpha \to 3 \, \ell_\beta$. These are $\gamma$-penguins,
$Z$-penguins, Higgs-penguins and box diagrams.  In our computations we
did not consider Higgs-penguins, since we are mostly interested in
processes involving the first two charged lepton generations, whose
small Yukawa couplings suppress Higgs contributions. Notice that this
assumption would not be valid for LFV processes involving $\tau$
leptons. However, the experimental limits in this case are not as
stringent as those found for processes involving the first two
generations, and thus their consideration would not change the
phenomenological picture.

\begin{figure}[t]
\begin{center}
\includegraphics[scale=0.7]{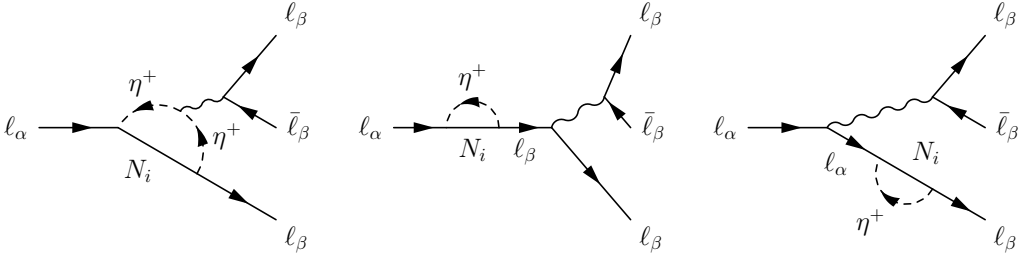}
\caption{Penguin contributions to
  $\ell_\alpha \to 3 \, \ell_\beta$. The wavy line
  represents either a photon or a Z-boson.}
\label{fig:m3e-g}
\end{center}
\end{figure}

\begin{figure}[t]
\begin{center}
\includegraphics[scale=0.7]{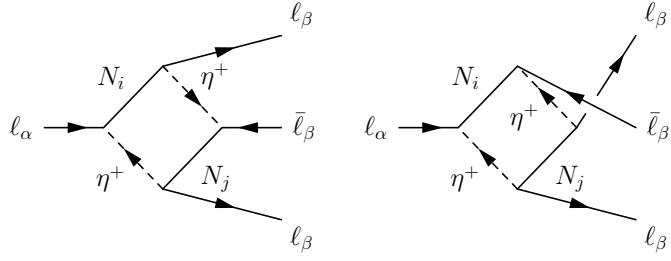}
\caption{Box contributions to
  $\ell_\alpha \to 3 \, \ell_\beta$.}
\label{fig:m3e-box}
\end{center}
\end{figure}

Let us consider the momentum assignment $\ell_\alpha(p) \to
\ell_\beta(k_1) \bar \ell_\beta(k_2) \ell_\beta(k_3)$.  Then, the
$\gamma$-penguin diagrams shown in Fig.~\ref{fig:m3e-g} lead to the
amplitude\footnote{In the presentation of our results we will follow a
  notation inspired by \cite{Arganda:2005ji}, which improved on
  \cite{Hisano:1995cp}.}
\begin{eqnarray}
i\mathcal{M}_{\gamma}&=&
ie^2 A_{ND} \, \bar{u}(k_1)\gamma^{\mu}P_Lu(p)\bar{u}(k_3)\gamma_{\mu}v(k_2)\nonumber\\
&&+ie^2\frac{m_\alpha}{q^2} A_D \, \bar{u}(k_1)\sigma^{\mu\nu}q_{\nu}P_Ru(p)
\bar{u}(k_3)\gamma_{\mu}v(k_2)
-(k_1\leftrightarrow k_3), \label{eq:photon-penguin}
\end{eqnarray}
where $q\equiv k_1-p$ is the photon momentum. Other operators turn out
to be suppressed by charged lepton masses and thus they are neglected
in Eq.~\eqref{eq:photon-penguin}. The coefficient $A_D$ was given in
Eq.~\eqref{eq:A2R}, whereas the coefficient $A_{ND}$, which
corresponds to the photonic non-dipole contributions, is given by
\begin{equation}
A_{ND}=\sum_{i=1}^3\frac{y_{i\beta}^*y_{i\alpha}}
{6(4\pi)^2}\frac{1}{m_{\eta^+}^2}
G_2\left(\xi_i\right), \label{eq:A1L}
\end{equation}
where the loop function $G_2(x)$ is given in appendix
\ref{sec:appendix1}.

Similarly, we now consider the contributions from $Z$-penguin
diagrams, also shown in Fig.~\ref{fig:m3e-g}. Neglecting sub-dominant
terms proportional to $q^2$, $q$ being the 4-momentum of the
$Z$-boson, the resulting amplitude can be written as
\begin{equation}
i\mathcal{M}_{Z}= \frac{iF}{m_Z^2} \, \bar{u}(k_1)\gamma^{\mu} P_R
u(p)\bar{u}(k_3) \gamma_{\mu} \left( g_L^\ell P_L + g_R^\ell P_R \right)
v(k_2) -(k_1\leftrightarrow k_3) \, ,
\end{equation}
where
\begin{equation}
g_L^\ell = \frac{g_2}{\cos \theta_W}\left( \frac{1}{2} - \sin^2 \theta_W
\right), \qquad g_R^\ell = - \frac{g_2}{\cos \theta_W} \sin^2 \theta_W,
\end{equation}
are the tree-level $Z$-boson couplings to a pair of charged
leptons. Here $g_2$ is the $SU(2)_L$ gauge coupling and $\theta_W$ is
the weak mixing angle. The coefficient $F$ is given by
\begin{equation}
F = \sum_{i=1}^3\frac{y_{i\beta}^*y_{i\alpha}}
{2(4\pi)^2}\frac{m_\alpha m_\beta}{m_{\eta^+}^2} \frac{g_2}{\cos \theta_W}
F_2\left(\xi_i\right) \, .
\label{eq:FR}
\end{equation}
Equation \eqref{eq:FR} shows that $Z$-penguins are suppressed by the
charged lepton masses $m_\alpha$ and $m_{\beta}$. Therefore, although
we fully derived and included them in our computation, we found that
they always have negligible contributions to the LFV processes
considered in this paper. For this reason, the total decay width for
$\ell_\alpha \to 3 \, \ell_\beta$ will be mainly
given by the $\gamma$-penguins and the box contributions, whose
relative size will determine the phenomenology.

Finally, the box diagrams contributing to the process
$\ell_\alpha \to 3 \, \ell_\beta$ are shown in
Fig.~\ref{fig:m3e-box}. One finds the following amplitude
\begin{eqnarray}
i \mathcal{M}_{\mathrm{box}} = ie^2B \left[ \bar u (k_3) \gamma^\mu P_L v(k_2) \right] \left[ \bar u (k_1) \gamma_\mu P_L u(p) \right].
\end{eqnarray}
The coefficient $B$ is given by\footnote{In \cite{Arganda:2005ji} this
  coefficient was denoted as $B_1^L$. The rest of box contributions
  are clearly suppressed in the scotogenic model.}
\begin{eqnarray}
e^2 B = \frac{1}{(4\pi)^2m_{\eta^+}^2} 
 \sum_{i,\:j=1}^3\left[ \frac{1}{2} D_1(\xi_i,\xi_j) y_{j \beta}^* y_{j \beta}
	   y_{i \beta}^* y_{i \alpha} + \sqrt{\xi_i\xi_j}
	   D_2(\xi_i,\xi_j) y_{j \beta}^* y_{j \beta}^* y_{i \beta}
	   y_{i \alpha}  \right],
\end{eqnarray}
where the loop functions $D_1(x,y)$ and $D_2(x,y)$ are given in
appendix \ref{sec:appendix1}. 
The branching ratio for
$\ell_\alpha \to 3 \, \ell_\beta$ is given by
\begin{eqnarray}
\text{Br}\left(\ell_{\alpha}\to
\ell_{\beta}\overline{\ell_{\beta}}\ell_{\beta}\right)&=&
\frac{3(4\pi)^2\alpha_{\mathrm{em}}^2}{8G_F^2}
\left[|A_{ND}|^2
  +|A_D|^2\left(\frac{16}{3}\log\left(\frac{m_\alpha}{m_\beta}\right)
  -\frac{22}{3}\right)+\frac{1}{6}|B|^2\right.\nonumber\\
 &&\left.+ \frac{1}{3} \left( 2  |F_{RR}|^2 + |F_{RL}|^2 \right)
    +\left(-2 A_{ND} A_D^{*}+\frac{1}{3}A_{ND} B^*
  -\frac{2}{3}A_D B^*+\mathrm{h.c.}\right)\right]\nonumber\\
&&\times \, \mathrm{Br}\left(\ell_{\alpha}\to\ell_{\beta}\nu_{\alpha}
\overline{\nu_{\beta}}\right) \, , \label{eq:l3lBR}
\end{eqnarray}
where $F_{RR}$ and $F_{RL}$ are given by
\begin{equation}
F_{RR} = \frac{F \, g_R^\ell}{g_2^2 \sin^2 \theta_W m_Z^2} \qquad ,
\qquad F_{RL} = \frac{F \, g_L^\ell}{g_2^2 \sin^2 \theta_W m_Z^2} \quad .
\end{equation}
In Eq. \eqref{eq:l3lBR}, the mass of the charged lepton in the final
state, $m_\beta$, is kept only in the logarithmic term, where it plays
the role of regulating the infrared divergence that would appear
otherwise.

\subsection{$\mu-e$ conversion in nuclei}

The most remarkable experimental projects in the near future will be
devoted to searches for $\mu-e$ conversion in nuclei. The great
sensitivities announced by the different collaborations might make
this observable the most stringent one in most neutrino mass models.
We will present our results using the notation and conventions of
Refs. \cite{Kuno:1999jp,Arganda:2007jw}. The conversion rate, relative
to the the muon capture rate, can be expressed as
\begin{align}
{\rm CR} (\mu- e, {\rm Nucleus}) &= 
\frac{p_e \, E_e \, m_\mu^3 \, G_F^2 \, \alpha_{\mathrm{em}}^3 
\, Z_{\rm eff}^4 \, F_p^2}{8 \, \pi^2 \, Z}  \nonumber \\
&\times \left\{ \left| (Z + N) \left( g_{LV}^{(0)} + g_{LS}^{(0)} \right) + 
(Z - N) \left( g_{LV}^{(1)} + g_{LS}^{(1)} \right) \right|^2 + 
\right. \nonumber \\
& \ \ \ 
 \ \left. \,\, \left| (Z + N) \left( g_{RV}^{(0)} + g_{RS}^{(0)} \right) + 
(Z - N) \left( g_{RV}^{(1)} + g_{RS}^{(1)} \right) \right|^2 \right\} 
\frac{1}{\Gamma_{\rm capt}}\,.
\end{align}
Here $Z$ and $N$ are the number of protons and neutrons in the
nucleus, $Z_{\rm eff}$ is the effective atomic charge (see
\cite{Chiang:1993xz}), $F_p$ is the nuclear matrix element and
$\Gamma_{\rm capt}$ represents the total muon capture rate. The values
of these parameters for the nuclei used in experiments can be found in
\cite{Arganda:2007jw} and references therein. Furthermore, $p_e$ and
$E_e$ (taken to be $\simeq m_\mu$ in the numerical evaluation) are the
momentum and energy of the electron and $m_\mu$ is the muon mass.  In
the above, $g_{XK}^{(0)}$ and $g_{XK}^{(1)}$ (with $X = L, R$ and $K =
S, V$) are given by
\begin{align}
g_{XK}^{(0)} &= \frac{1}{2} \sum_{q = u,d,s} \left( g_{XK(q)} G_K^{(q,p)} +
g_{XK(q)} G_K^{(q,n)} \right)\,, \nonumber \\
g_{XK}^{(1)} &= \frac{1}{2} \sum_{q = u,d,s} \left( g_{XK(q)} G_K^{(q,p)} - 
g_{XK(q)} G_K^{(q,n)} \right)\,.
\end{align}
The numerical values of the $G_K$ coefficients can be found in
\cite{Kuno:1999jp,Kosmas:2001mv,Arganda:2007jw}.

As for $\ell_\alpha \to 3 \, \ell_\beta$, the
$\mu-e$ conversion rate receives contributions from $\gamma$-, $Z$-
and Higgs-penguins. Note, however, the absence of box contributions
(besides the tiny SM contribution). This is due to the unbroken
$\mathbb{Z}_2$ symmetry, which forbids the coupling between the
$\eta^\pm$ scalars and the quark sector. Moreover, we neglect again
the Higgs-penguin contributions due to the smallness of the involved
Yukawa couplings. Therefore, the corresponding couplings are
\begin{eqnarray}
g_{LV(q)} &=& g_{LV(q)}^{\gamma} + g_{LV(q)}^{Z}\,, \nonumber \\
g_{RV(q)} &=& \left. g_{LV(q)} \right|_{L \leftrightarrow R}\,, \nonumber \\
g_{LS(q)} &\approx& 0 \, , \nonumber \\ 
g_{RS(q)} &\approx& 0 \, .
\end{eqnarray}
\begin{figure}[t]
\begin{center}
\includegraphics[scale=0.7]{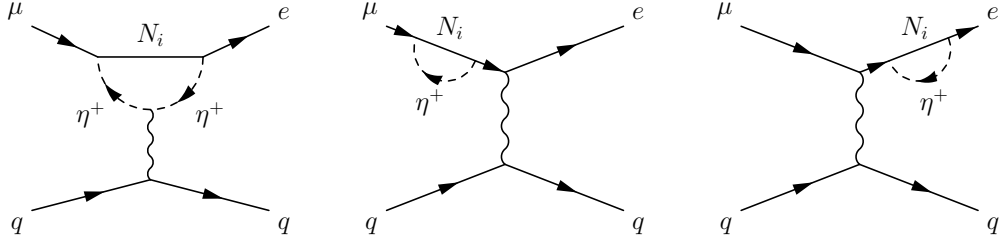}
\caption{Penguin contributions to $\mu-e$ conversion in nuclei. The
  wavy line represents either a photon or a Z-boson.}
\label{fig:mueconversion}
\end{center}
\end{figure}
The photon and $Z$-boson couplings can be computed from the Feynman
diagrams in Fig.~\ref{fig:mueconversion}. One finds that the relevant
(non-negligible) couplings are
\begin{align}
g_{LV(q)}^{\gamma} &= \frac{\sqrt{2}}{G_F} e^2 Q_q 
\left(A_{ND} - A_D \right)\,, \nonumber \\
g_{RV(q)}^{Z} &= -\frac{\sqrt{2}}{G_F} \, \frac{g_L^q + g_R^q}{2} \, 
\frac{F}{m_Z^2} \,. 
\end{align}
The form factors $A_{ND}$, $A_D$ and $F$ are given in section
\ref{subsec:l3l}, see equations \eqref{eq:A1L}, \eqref{eq:A2R} and
\eqref{eq:FR}. Furthermore, $Q_q$ is the electric charge of the
corresponding quark and
\begin{equation}
g_L^q = \frac{g_2}{\cos \theta_W}\left( Q_q \sin^2 \theta_W - T_3^q
\right), \qquad g_R^q = \frac{g_2}{\cos \theta_W} Q_q \sin^2 \theta_W,
\end{equation}
are the tree-level $Z$-boson couplings to a pair of quarks.

\section{Phenomenological discussion}
\label{sec:pheno}

In this section we present and discuss our numerical results. We will
explore the parameter space and highlight some relevant
phenomenological issues which, to the best of our knowledge, have not
been discussed in the existing literature.

In the numerical evaluation of our results we considered both
hierarchies for the light neutrino spectrum~\footnote{In our
  conventions, the lightest neutrino mass is $m_1$ for normal
  hierarchy and $m_3$ for inverted hierarchy, although we will denote
  it by $m_{\nu_1}$ in general.}, normal hierarchy (NH) and inverted
hierarchy (IH), and randomly chose the neutrino oscillation parameters
in the $1 \sigma$ ranges found by the global fit
\cite{GonzalezGarcia:2012sz} (\textit{Free Fluxes + RSBL} results). We
note that these ranges are in good agreement with the ones found by
other fits, see Refs.~\cite{Tortola:2012te,Fogli:2012ua}. For
$\theta_{23}$, the atmospheric angle, we selected the local minimum in
the first octant, in agreement with \cite{Fogli:2012ua}.

Unless explicitly expressed otherwise, all our numerical results were
obtained for a degenerate right-handed neutrino spectrum, assuming a
random real $R$ matrix and $\lambda_5 = 10^{-9}$. This value was found
in \cite{Schmidt:2012yg} to be compatible with a correct right-handed
neutrino DM relic density due to the resulting size of the Yukawa
couplings. Moreover, note that it is natural for $\lambda_5$ to be
very small since, in case it was exactly zero, a definition of a
conserved lepton number would be possible \cite{Kubo:2006yx}.

\subsection{The ratio $\text{Br}(\ell_\alpha \to 3 \, \ell_\beta)/\text{Br}(\ell_\alpha \to \ell_\beta \gamma)$}
\label{subsec:R}

Most LFV phenomenological studies focus on the radiative decay $\mu
\to e \gamma$, ignoring other LFV observables. There are two reasons
for this. First, the great performance of the MEG experiment, that
recently set the quite impressive bound $\text{Br}(\mu \to e \gamma) <
5.7 \times 10^{-13}$. And second, the dipole dominance in many models
of interest. When the dipole contributions originated in photon
penguin diagrams dominate, the rate for $\mu \to 3 e$ is correlated
with the rate for $\mu \to e \gamma$. In this case a simple relation
can be derived \cite{Arganda:2005ji}
\begin{equation}
\text{Br}(\mu \to 3 e) \simeq \frac{\alpha_{\mathrm{em}}}{3 \pi} 
\left(\log\left(\frac{m^2_{\mu}}{m^2_{e}}\right) - \frac{11}{4} \right) 
\text{Br}(\mu \to e \gamma) \, .
\end{equation}
Since the proportionality factor is much smaller than one, $\mu \to 3
e$ is suppressed with respect to $\mu \to e \gamma$ and the latter
becomes the process leading to the most stringent constraints. This
assumption has been present in all previous works on lepton flavor
violation in the scotogenic model~\cite{Kubo:2006yx, Sierra:2008wj,
  Suematsu:2009ww, Adulpravitchai:2009gi}. They have either assumed
explicitly that photon penguin diagrams dominate or simply ignored
4-fermion observables (like $\mu \to 3 e$) and concentrated on $\mu
\to e \gamma$ (an approach consistent with the assumption that the
photonic dipole contributions dominate).
Here we want to study under what conditions that is a bad
simplification of the phenomenology. In order to do so, we consider
the ratio\footnote{We concentrate here on $\mu$ decays due to the
  better experimental bounds and perspectives. Similar results are
  obtained for $\tau$ decays.}
\begin{equation}
\label{eq:ratioR}
R_{\mu e} = \frac{\text{Br}(\mu \to 3 e)}{\text{Br}(\mu \to e \gamma)} \, .
\end{equation}
In those regions of parameter space where $R_{\mu e} > 1$, the observable that
provides the most stringent limits is $\text{Br}(\mu \to 3 e)$,
whereas $\text{Br}(\mu \to e \gamma)$ would be the most relevant
observable in regions where $R_{\mu e} < 1$.

\begin{figure}[t]
\centering \includegraphics[width=0.6\linewidth]{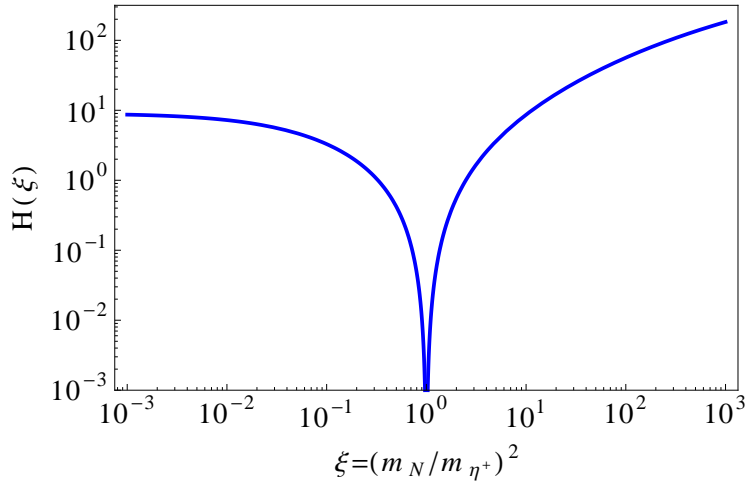}
\caption{$H(\xi)$ as a function of $\xi = (m_N/m_{\eta^+})^2$. For the
  definition see Eq. \eqref{eq:Hxi}.}
\label{fig:Hxi}
\end{figure}

Since the photonic dipole operators contribute to both observables,
the only way to obtain $R_{\mu e} > 1$ is to have dominant
contributions from box and/or photonic non-dipole diagrams in $\mu \to
3 e$ ($Z$-penguins are suppressed by charged leptons and thus their
contribution is always negligible). Since the photonic non-dipole
diagrams, given by the $A_{ND}$ form factor, never exceed the dipole
ones as much as to compensate the large factor that multiplies
$|A_D|^2$ in the branching ratio formula (see Eq. \eqref{eq:l3lBR}),
they are never dominant. We are therefore left with a
\textit{competition} between photonic dipole operators and box
diagrams.

Assuming box dominance in $\mu \to 3 e$ and a degenerate right-handed
neutrino spectrum one can estimate
\begin{equation}
\label{eq:ratioRapp}
R_{\mu e} \sim \frac{y^4}{48 \pi^2 e^2} H(\xi),
\end{equation}
where $y$ is the average size of the Yukawa coupling and the function
$H(\xi)$ is defined as
\begin{equation}
\label{eq:Hxi}
H(\xi) = \left( \frac{\frac{1}{2} D_1(\xi,\xi) + \xi D_2(\xi,\xi)}{F_2(\xi)} \right)^2 \, .
\end{equation}
The function $H(\xi)$ is shown in Fig. \ref{fig:Hxi}. Notice the
cancellation for $\xi = 1$. This pole is caused by an exact
cancellation between the contributions from the loop functions $D_1$
and $D_2$. However, for $\xi \ll 1$ and $\xi \gg 1$ one always has
$H(\xi) > 1$.

\begin{figure}[t]
\centering
\includegraphics[width=0.49\linewidth]{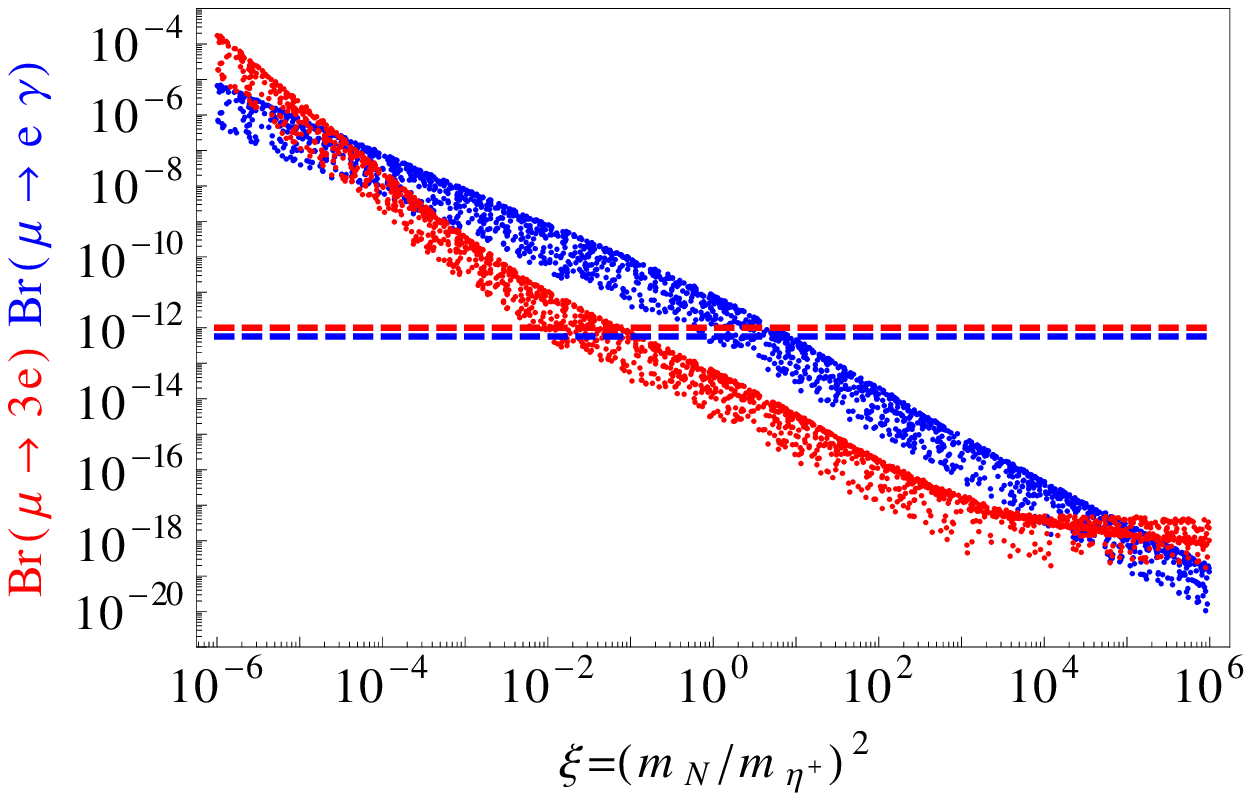}
\includegraphics[width=0.49\linewidth]{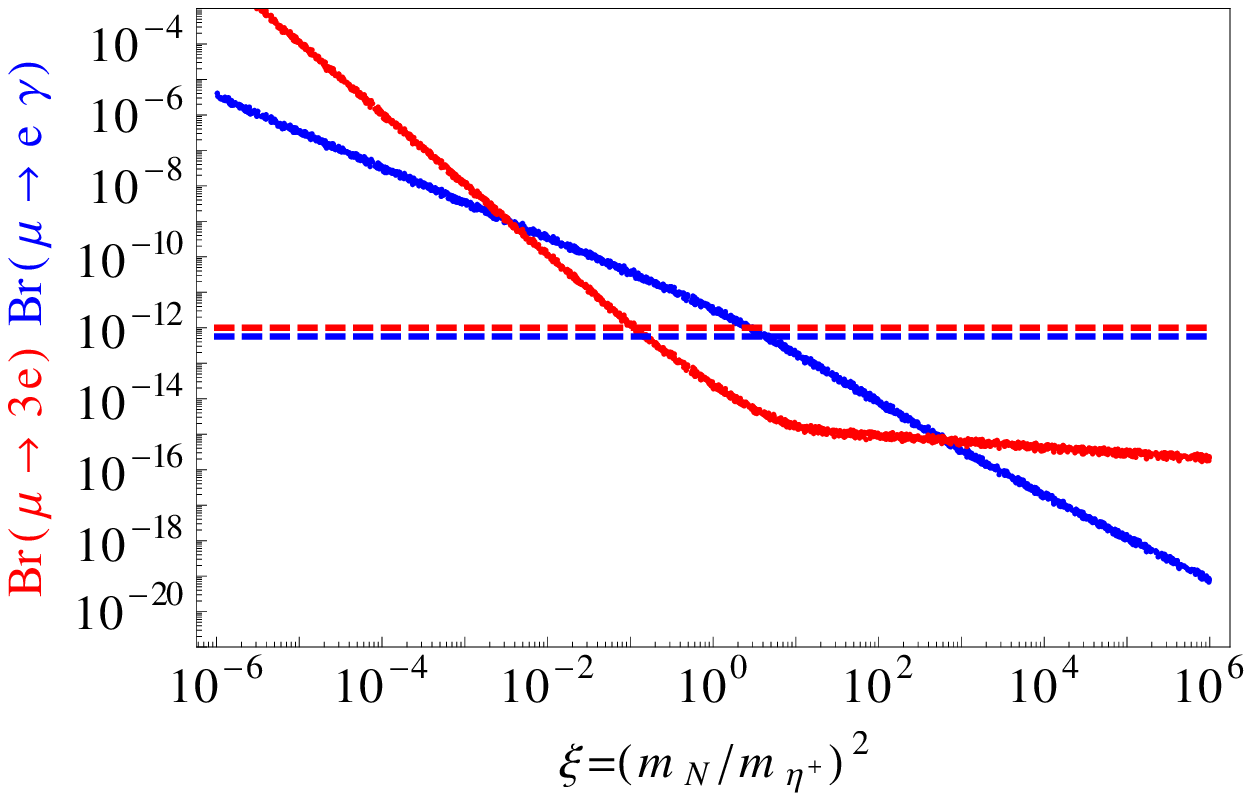}
\caption{$\text{Br}(\mu \to e \gamma)$ and $\text{Br}(\mu \to 3 e)$ as
  a function of $\xi = (m_N/m_{\eta^+})^2$. A degenerate right-handed
  neutrino spectrum has been assumed, see text for details. To the
  left for NH, whereas to the right for IH. The horizontal dashed
  lines show the current upper bounds.}
\label{fig:BRs-xi}
\end{figure}

It is clear from Eq. \eqref{eq:ratioRapp} and Fig. \ref{fig:Hxi} that
in order to increase the value of $R_{\mu e}$ one requires large
Yukawa couplings and a large mass difference between the right-handed
neutrinos and the $\eta$ scalars (in order to be far from $\xi =
1$). This is illustrated in Fig. \ref{fig:BRs-xi}, where we show
$\text{Br}(\mu \to e \gamma)$ (blue) and $\text{Br}(\mu \to 3 e)$
(red) as a function of $\xi = (m_N/m_{\eta^+})^2$. The horizontal
dashed lines represent the current upper bounds on the branching
ratios. Fixed values $m_{\eta^+} = 1$ TeV and $m_{\nu_1} = 10^{-3}$ eV
(lightest neutrino mass) are taken. On the left-hand side we show our
results for NH, whereas the right-hand side shows our results for
IH. A random Dirac phase $\delta$ has been taken. As can be derived
from the spread of the points, this parameter has a much larger
influence for NH.
As expected from our previous estimate, one can in principle have
$R_{\mu e} > 1$ (or equivalently, $\text{Br}(\mu \to 3 e) >
\text{Br}(\mu \to e \gamma)$) for $\xi$ values far from $1$. Although
the region with $m_N \ll m_{\eta^+}$ is already excluded for this
value of $\lambda_5$, the region with $m_N \gg m_{\eta^+}$ is
compatible with all experimental constraints. Note that in this figure
all points have $\mathcal{O}(1)$ Yukawa couplings. Larger values for
$\lambda_5$ would decrease the size of the Yukawa couplings (see
Eq. \eqref{eq:nu-mass}), which in turn would imply a reduction of all
LFV rates.

Another parameter that turns out to be very relevant in the
determination of the ratio $R_{\mu e}$ is $m_{\nu_1}$, the mass of the
lightest neutrino. In order to illustrate this fact, we consider two
scenarios: (i) Scenario A: $m_N = 1$ TeV and $m_{\eta^+} = 4$ TeV, and
(ii) Scenario B: $m_N = 4$ TeV and $m_{\eta^+} = 1$ TeV. In both cases
we assume a degenerate right-handed neutrino spectrum, a random Dirac
phase and a random real $R$ matrix.

\begin{figure}[t]
\centering
\includegraphics[width=0.49\linewidth]{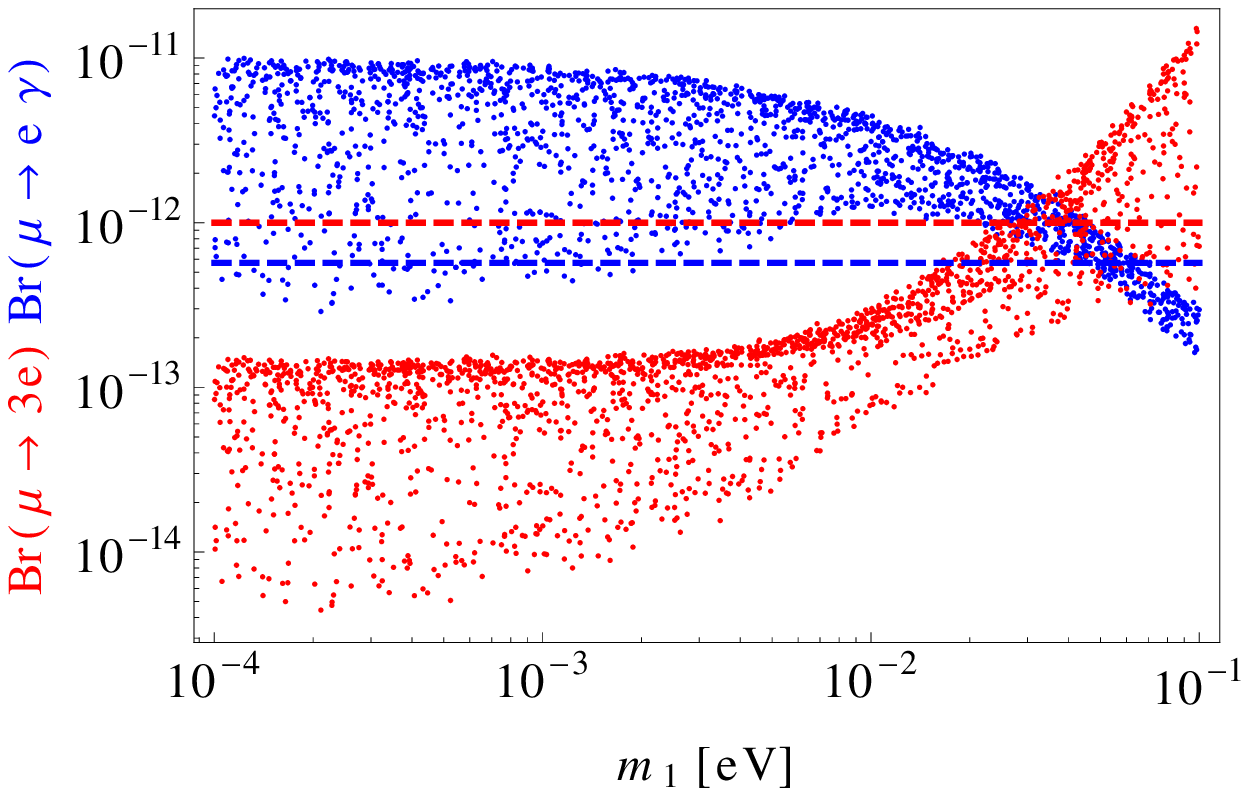}
\includegraphics[width=0.49\linewidth]{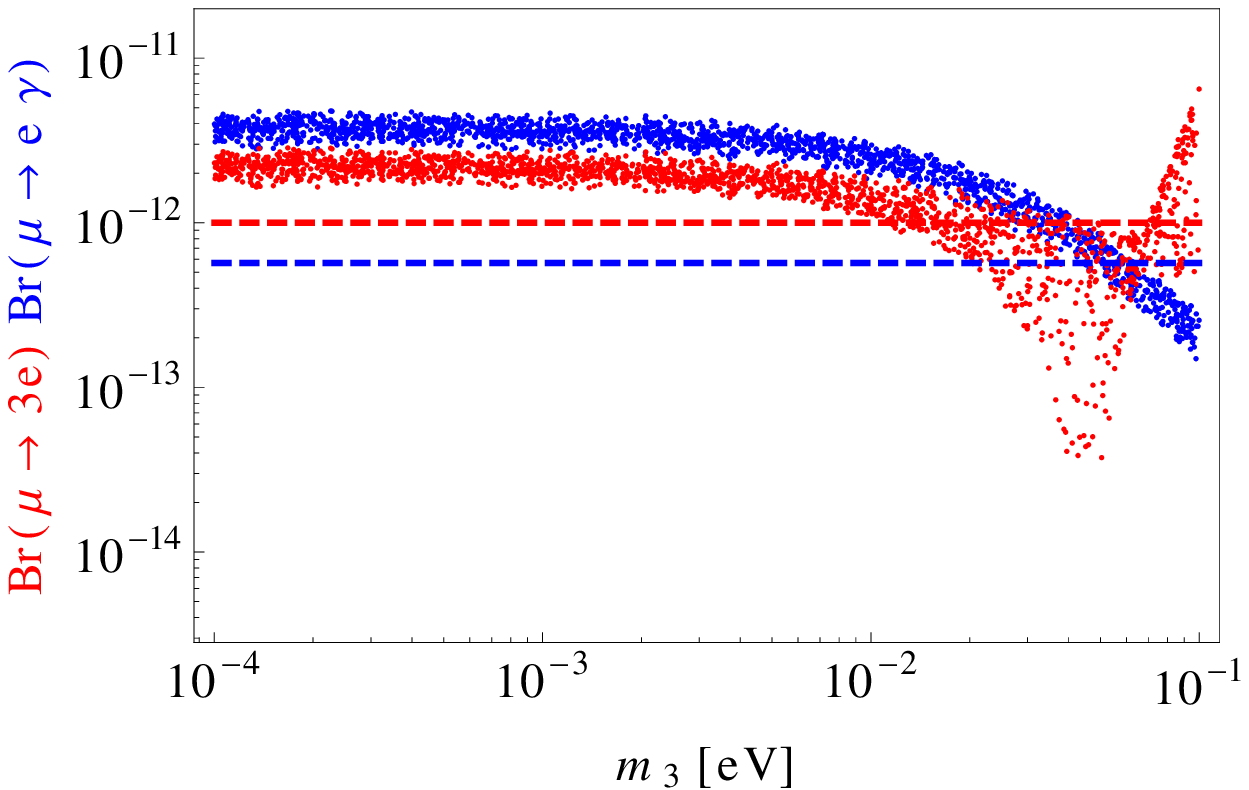}
\caption{$\text{Br}(\mu \to e \gamma)$ and $\text{Br}(\mu \to 3 e)$ as
  a function of the lightest neutrino mass. Scenario A is assumed, see
  text for details. To the left for NH, whereas to the right for
  IH. The horizontal dashed lines show the current upper bounds.}
\label{fig:BRs-m1-A}
\end{figure}

\begin{figure}[t]
\centering
\includegraphics[width=0.49\linewidth]{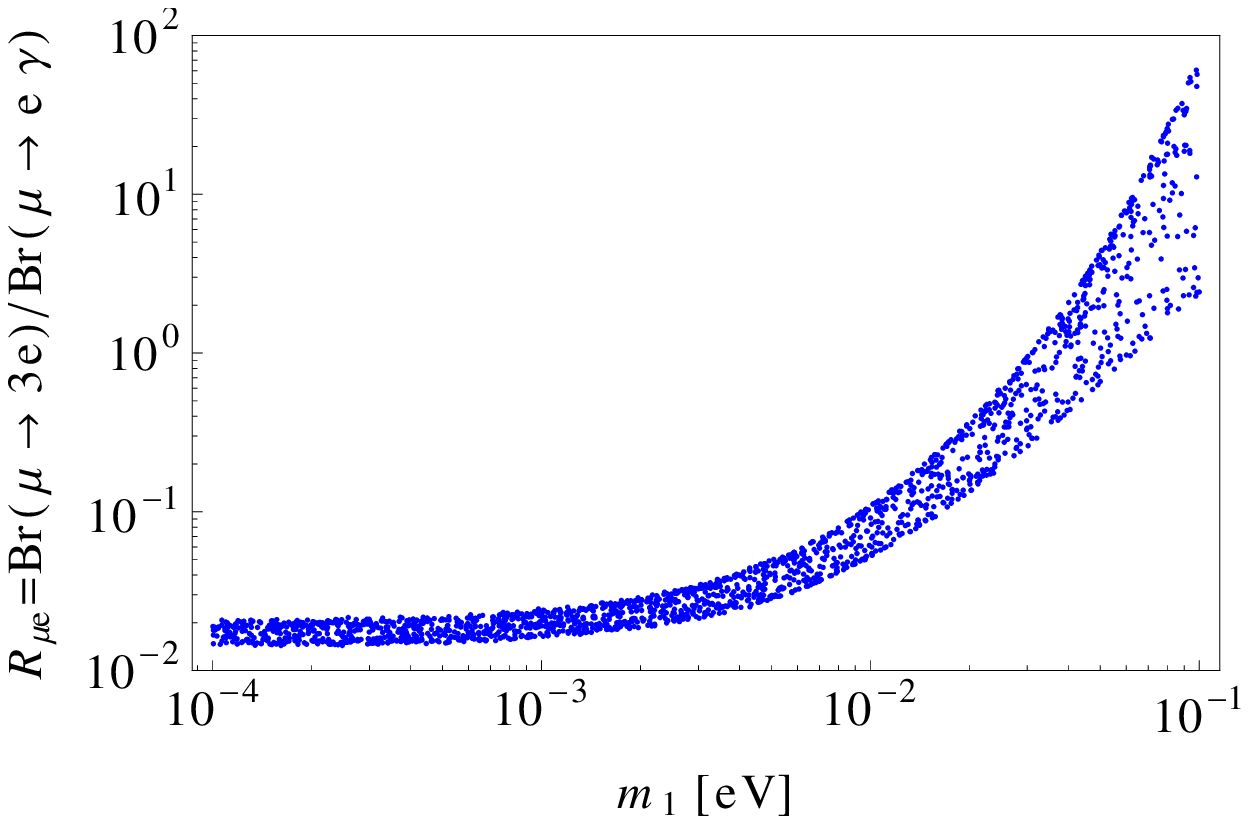}
\includegraphics[width=0.49\linewidth]{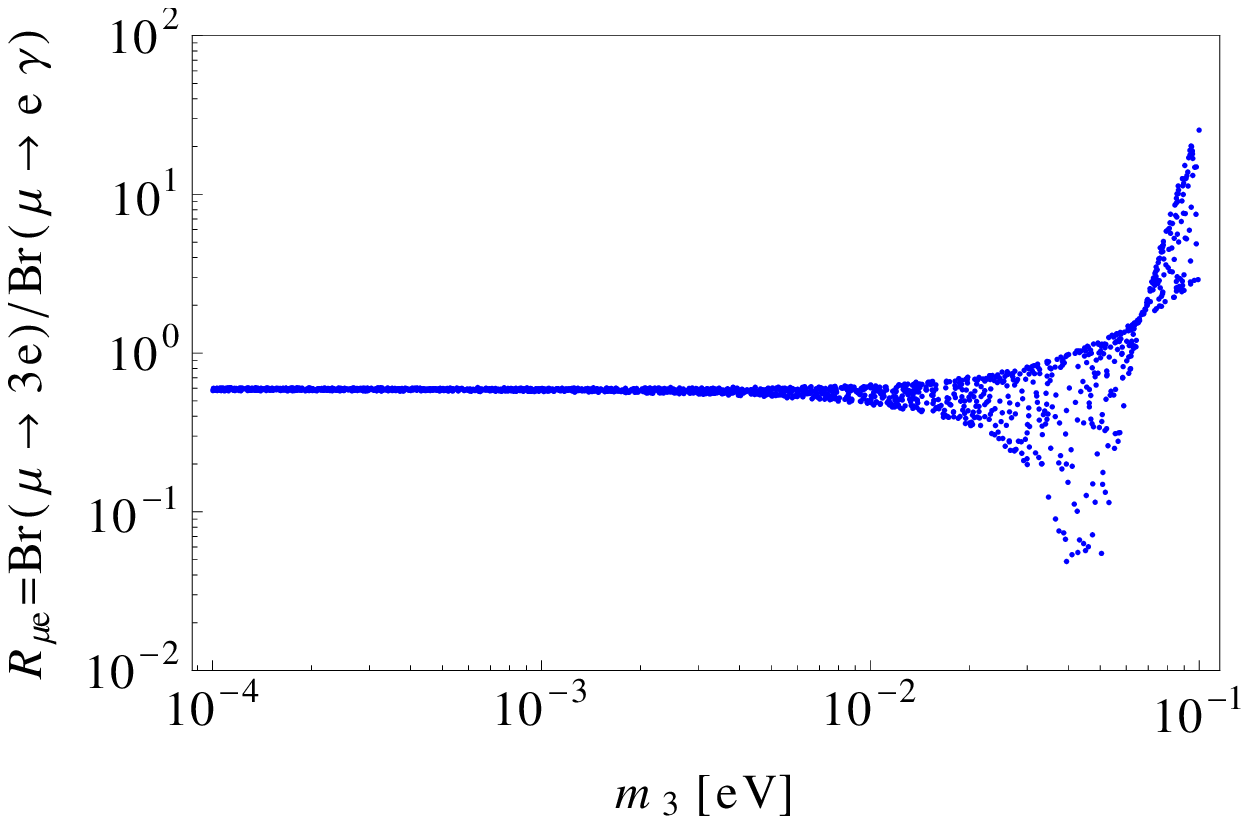}
\caption{The ratio $R_{\mu e} = \text{Br}(\mu \to 3 e)/\text{Br}(\mu \to e
  \gamma)$ as a function of the lightest neutrino mass. Scenario A is
  assumed, see text for details. To the left for NH, whereas to the
  right for IH.}
\label{fig:R-m1-A}
\end{figure}

Our numerical results for scenario A are shown in
Figs. \ref{fig:BRs-m1-A} and \ref{fig:R-m1-A}. The left-hand side of
these figures were obtained with NH, whereas the right-hand side shows
our results for IH. We see that large values of the lightest neutrino
mass may lead to large variations in the $\text{Br}(\mu \to e \gamma)$
and $\text{Br}(\mu \to 3 e)$ branching ratios, and thus in the ratio
$R_{\mu e}$. We conclude that the LFV rates in the scotogenic model
are very sensitive to the absolute scale for neutrino masses.

Figure \ref{fig:BRs-m1-A} demonstrates that in scenario A the neutrino
mass hierarchy also has a clear impact on the LFV rates. While for low
$m_{\nu_1}$, $\text{Br}(\mu \to 3 e)$ is clearly below the upper bound
for NH, it is largely excluded for IH since it exceeds it. Similarly,
while for low $m_{\nu_1}$ the ratio $R_{\mu e}$ is $\sim 10^{-2}$ in
NH (as expected from dipole domination), the contributions from box
diagrams already lead to a small increase for IH, where $R_{\mu e}
\sim 0.5$.

These figures can be understood by analyzing how the Yukawa couplings
depend on $m_{\nu_1}$. In particular, we must study the combinations
of Yukawa couplings that contribute to the LFV processes considered
here. Let us suppose that box diagrams dominate $\ell_\alpha \to 3 \,
\ell_\beta$ (otherwise we would be in a dipole dominated scenario
where the ratio $R_{\mu e}$ would not deviate significantly from $\sim
10^{-2}$). Then we have the relations
\begin{eqnarray}
\text{Br}(\ell_\alpha \to \ell_\beta \gamma) &\propto&
 \left| \left(y^\dagger y \right)_{\beta\alpha} \right|^2 \label{LLG-yy} \\ 
\text{Br}(\ell_\alpha \to 3 \, \ell_\beta) &\propto& 
\left| \frac{1}{2} D_1(\xi,\xi) \left( y^\dagger y \right)_{\beta \beta} \left( y^\dagger y \right)_{\beta \alpha} + \xi D_2(\xi,\xi) \left( y^T y \right)_{\beta \beta} \left( y^T y \right)_{\beta \alpha} \right|^2 \, . \label{L3L-yy}
\end{eqnarray}
Assuming degenerate right-handed neutrinos and a real $R$ matrix, we
can use the Casas-Ibarra parametrization in
Eq.~\eqref{eq:casas-ibarra} to obtain
\begin{eqnarray}
y^\dagger y &\propto& U_{\mathrm{PMNS}} \, \hat{m}_\nu \, U_{\mathrm{PMNS}}^{\dag} \label{ydagy} \\  
y^T y &\propto& U_{\mathrm{PMNS}}^* \, \hat m_\nu \, U_{\mathrm{PMNS}}^{\dag} \, . \label{yTy}
\end{eqnarray}
Analytical results for the relevant elements of the matrix
combinations (or \textit{flavor structures}) in the previous
expressions can be found in appendix \ref{sec:appendix2}.

Let us first focus on the NH case. Notice that in scenario A we have
$\xi = (1/4)^2 = 0.0625$. With such a small value for $\xi$, we expect
the $D_1$ term in the box contribution to dominate over the $D_2$
term. Therefore, we must inspect the expressions for $\left( y^\dagger
y \right)_{21}$ and $\left( y^\dagger y \right)_{22}$ or, as shown
above, $\left( U_{\mathrm{PMNS}} \, \hat{m}_\nu \,
U_{\mathrm{PMNS}}^{\dag} \right)_{21}$ and $\left( U_{\mathrm{PMNS}}
\, \hat{m}_\nu \, U_{\mathrm{PMNS}}^{\dag} \right)_{22}$. On the one
hand, we see that $\left( U_{\mathrm{PMNS}} \, \hat{m}_\nu \,
U_{\mathrm{PMNS}}^{\dag} \right)_{21}$ depends only on differences of
mass eigenvalues. Therefore, it decreases for higher values of the
lightest neutrino mass. This can be easily understood from the
expansion $m_j - m_i = \Delta m_{ji}^2/(2m_i) + \dots$,
where $\Delta m_{ji}^2 = m_j^2 - m_i^2$ is the corresponding squared mass
difference. This expansion is valid for $\Delta m_{ji}^2/m_i^2\ll1$. On the other hand, Eq. \eqref{eq:dag22} clearly shows that
$\left( U_{\mathrm{PMNS}} \, \hat{m}_\nu \, U_{\mathrm{PMNS}}^{\dag}
\right)_{22}$ increases for higher values of the lightest neutrino
mass. This explains why $\text{Br}(\mu \to e \gamma)$ decreases with
$m_{\nu_1}$ while $\text{Br}(\mu \to 3 e)$ increases. The resulting
behavior for the ratio $R_{\mu e}$ is then trivially deduced from
these considerations. Notice that this quantity can reach values as
high as $\sim 50$. In this case it is obvious that one cannot ignore
$\text{Br}(\mu \to 3 e)$, but in fact this branching ratio becomes the
most relevant LFV observable.

The discussion for IH would be a bit more involved. In this case we
find a larger relevance of the $D_2$ piece. In fact, for $m_{\nu_1}
\sim 10^{-2}$ eV this term competes with the $D_1$ term, leading to
the feature observed on the right-hand sides of
Figs. \ref{fig:BRs-m1-A} and \ref{fig:R-m1-A}.

\begin{figure}[t]
\centering
\includegraphics[width=0.49\linewidth]{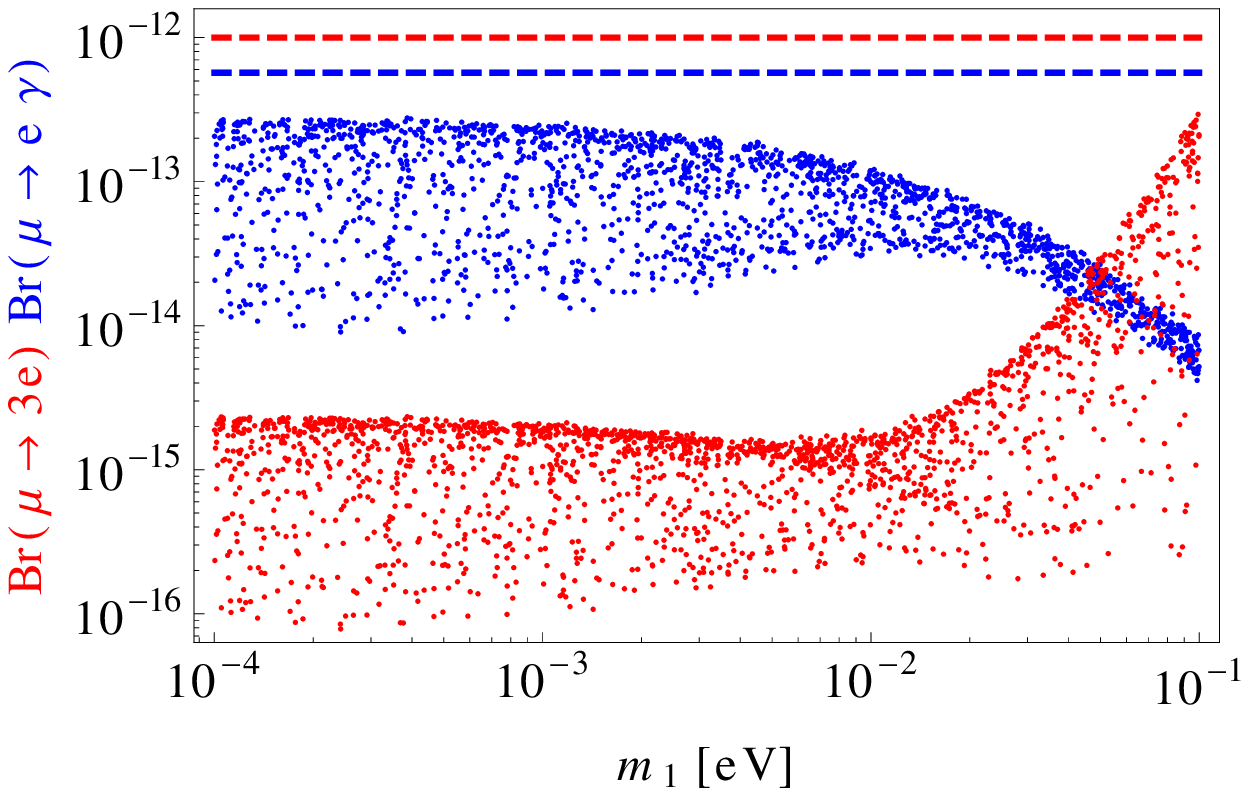}
\includegraphics[width=0.49\linewidth]{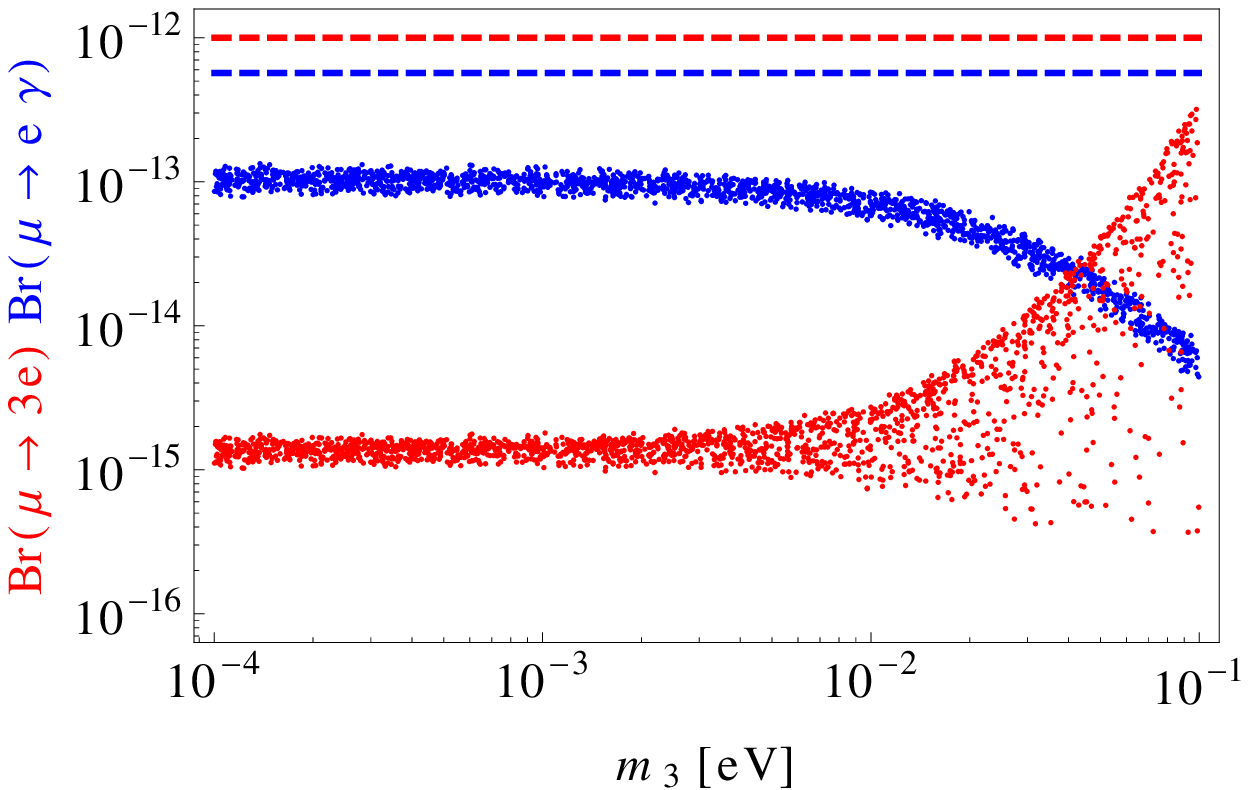}
\caption{$\text{Br}(\mu \to e \gamma)$ and $\text{Br}(\mu \to 3 e)$ as
  a function of the lightest neutrino mass. Scenario B is assumed, see
  text for details. To the left for NH, whereas to the right for
  IH. The horizontal dashed lines show the current upper bounds.}
\label{fig:BRs-m1-B}
\end{figure}

\begin{figure}[t]
\centering
\includegraphics[width=0.49\linewidth]{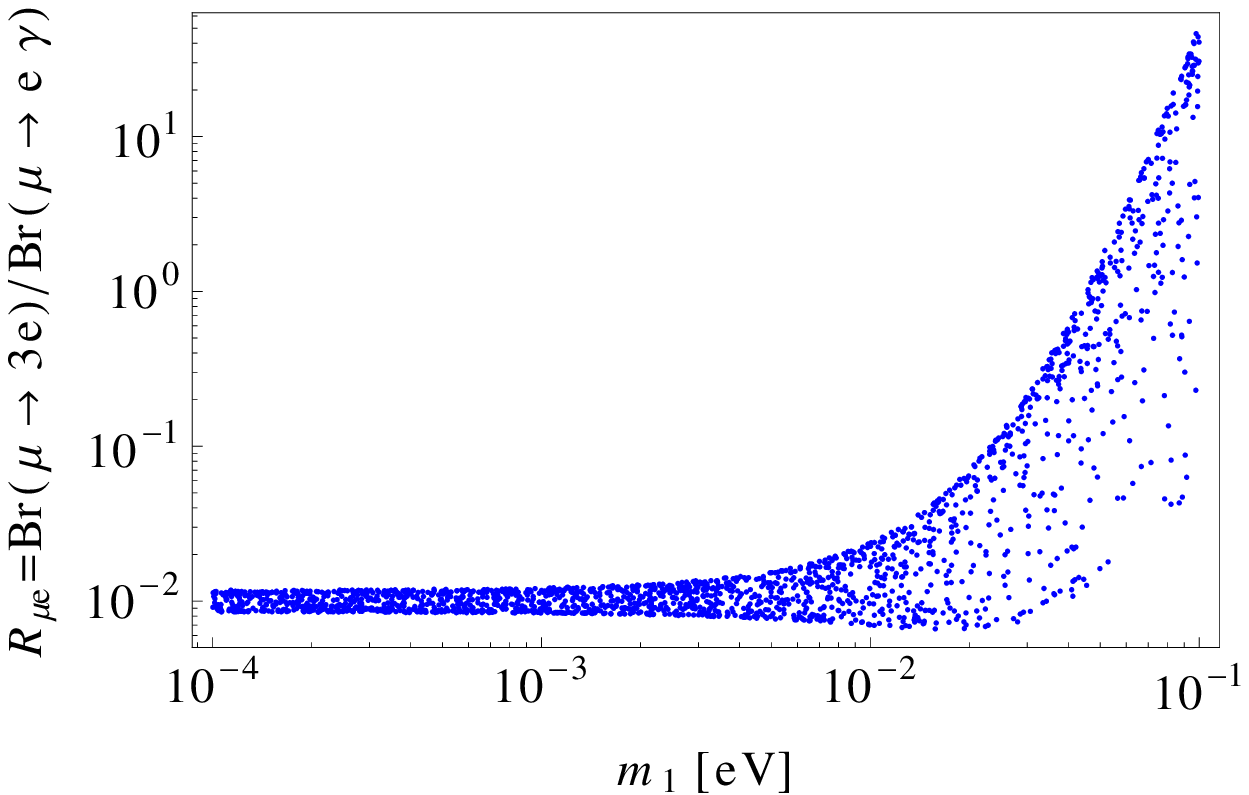}
\includegraphics[width=0.49\linewidth]{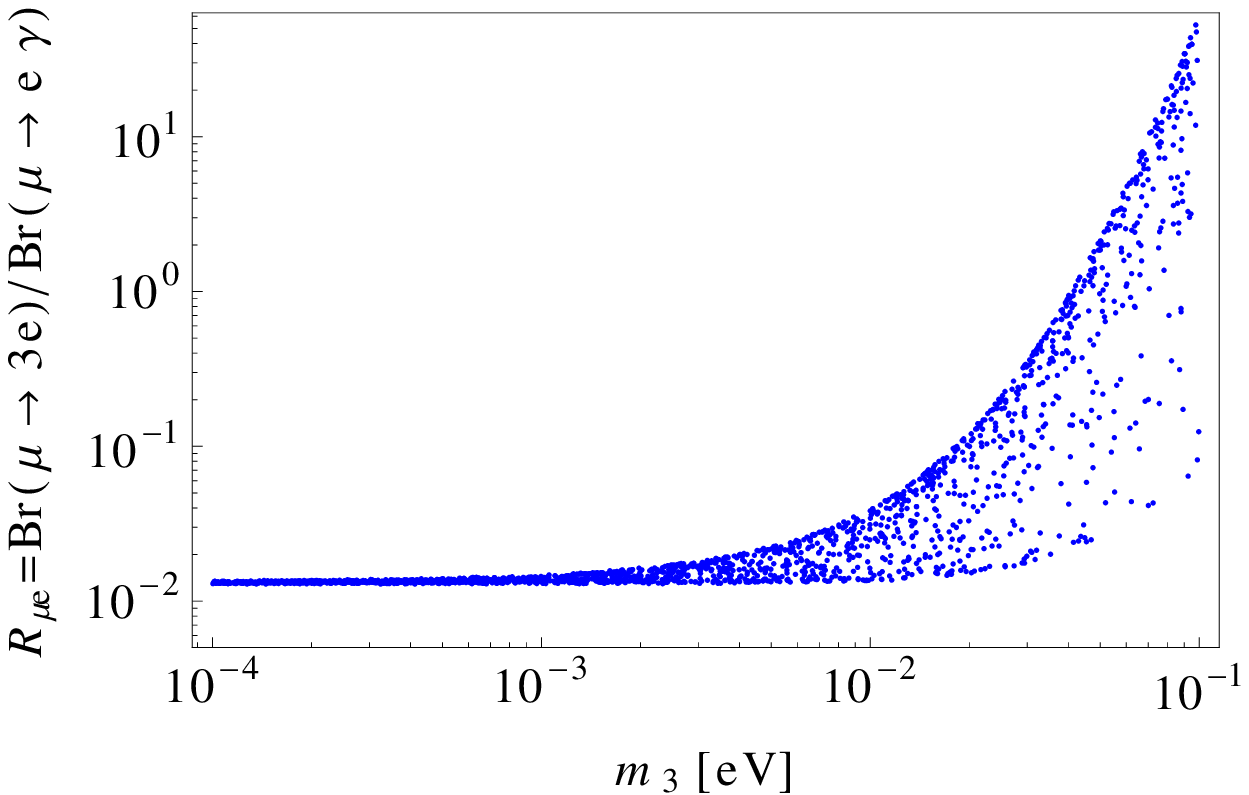}
\caption{The ratio $R_{\mu e} = \text{Br}(\mu \to 3 e)/\text{Br}(\mu \to e
  \gamma)$ as a function of the lightest neutrino mass. Scenario B is
  assumed, see text for details. To the left for NH, whereas to the
  right for IH.}
\label{fig:R-m1-B}
\end{figure}

Let us now consider our results for scenario B, shown in
Figs. \ref{fig:BRs-m1-B} and \ref{fig:R-m1-B}. Again, we present our
results for NH on the left-hand side and our results for IH on the
right-hand side. Regarding NH, it is already clear at first sight that
the results are qualitatively very similar to those found in scenario
A. Although the LFV rates are very different (much lower in this
case), the dependence on $m_{\nu_1}$ is very similar. Notice that all
points in these figures are actually allowed by the current
limits. This was expected, since it is well-known that LFV constraints
are more easily satisfied in scenarios with $m_N > m_{\eta^+}$
\cite{Kubo:2006yx}. On the other hand, the difference between NH and
IH found in scenario A is not present in scenario B, in which both
cases show the same behavior.

Finally, let us briefly discuss a scenario with non-degenerate
right-handed neutrinos. The spectrum in the right-handed neutrino
sector has an impact on the LFV rates, as we want to illustrate
here. In order to do so, we consider a spectrum of the type $m_N =
(\tilde{m}_N,\bar m_N^{(1)},\bar m_N^{(2)})$, with two fixed mass eigenvalues
($\bar m_N^{(1,2)}$) and one varying ($\tilde{m}_N$). Although one can
imagine other scenarios, this simple family of non-degenerate spectra
serves to show the qualitative behavior that we want to emphasize.

\begin{figure}[t]
\centering
\includegraphics[width=0.49\linewidth]{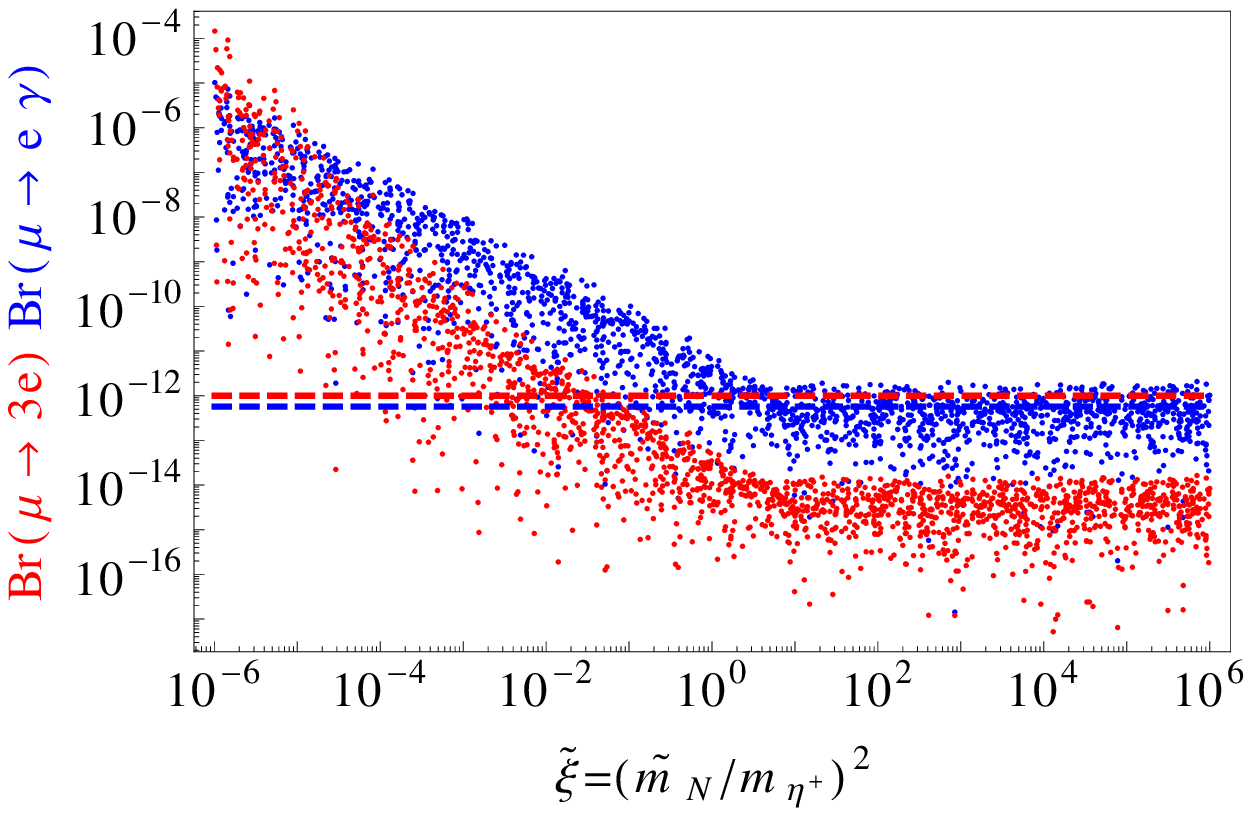}
\includegraphics[width=0.49\linewidth]{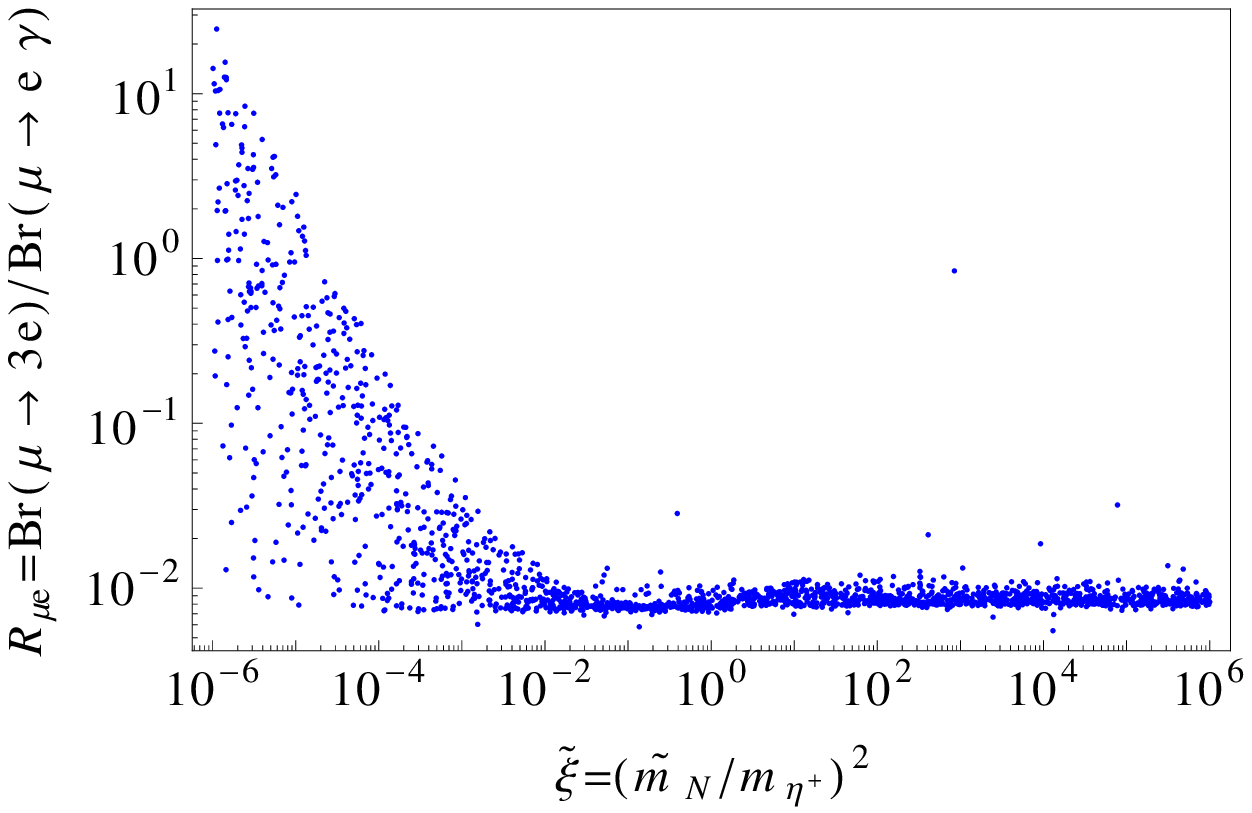}
\caption{$\text{Br}(\mu \to e \gamma)$ and $\text{Br}(\mu \to 3 e)$
  (to the left) and the resulting $R_{\mu e}$ ratio (to the right) as
  a function of $\tilde{\xi} = (\tilde{m}_N/m_{\eta^+})^2$. Normal
  hierarchy for the light neutrinos and a non-degenerate right-handed
  neutrino spectrum (with $\bar m_N^{(1)} = 2$ TeV and $\bar m_N^{(2)}
  = 3$ TeV) have been assumed, see text for details. The horizontal
  dashed lines show the current upper bounds.}
\label{fig:nondeg}
\end{figure}

Fig. \ref{fig:nondeg} shows a representative example of how the LFV
rates can change in a non-degenerate right-handed neutrino
spectrum. On the left, we show $\text{Br}(\mu \to e \gamma)$ (blue)
and $\text{Br}(\mu \to 3 e)$ (red) as a function of $\tilde{\xi} =
(\tilde{m}_N/m_{\eta^+})^2$, where the horizontal dashed lines
represent the current upper bounds on the branching ratios. On the
right we show the resulting $R_{\mu e}$ ratio. Fixed values $\bar
m_N^{(1)} = 2$ TeV, $\bar m_N^{(2)} = 3$ TeV, $m_{\eta^+} = 1$ TeV and
$m_{\nu_1} = 10^{-3}$ eV were assumed. A NH spectrum for the light
neutrinos was chosen for this figure and we allowed for a random Dirac
phase $\delta$.

As naively expected, low $\tilde{m}_N$ values enhance both branching
ratios, with $\text{Br}(\mu \to 3 e)$ being the one that typically
gets the larger enhancements. This is caused by the large box
contributions induced by the lightest right-handed neutrino. On the
other hand, when $\tilde{m}_N \gg \bar m_N^{(1,2)}$, the contribution
of the heaviest right-handed neutrino (with a mass $\tilde{m}_N$)
becomes sub-dominant and the LFV rates remain barely the same as in
the degenerate case. This implies that the general conclusions drawn
from the numerical results shown in this section are not restricted to
degenerate scenarios. Besides this fact, we do not find any other
remarkable feature in the LFV phenomenology for non-degenerate
right-handed neutrinos.

\subsection{Sensitivity to low-energy neutrino parameters}
\label{subsec:nuparam}

We have already shown the relevant role played by the lightest
neutrino mass in the resulting LFV branching ratios. Let us now extend
the discussion to the other undetermined low-energy parameter (besides
the Majorana phases), the Dirac phase $\delta$.

\begin{figure}[t]
\centering
\includegraphics[width=0.49\linewidth]{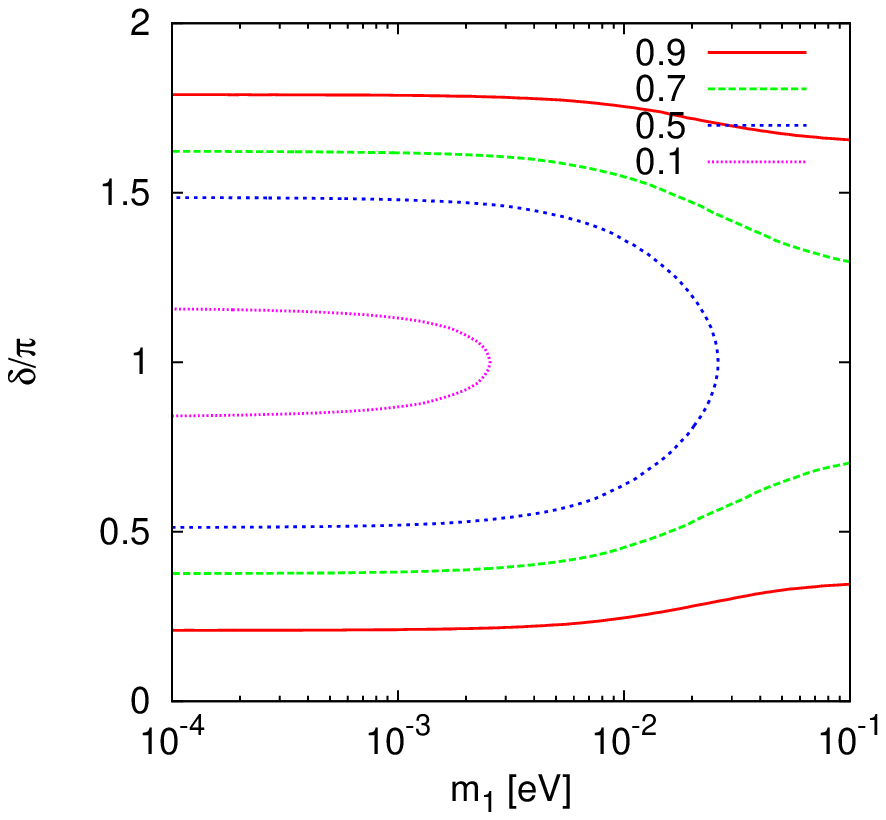}
\includegraphics[width=0.49\linewidth]{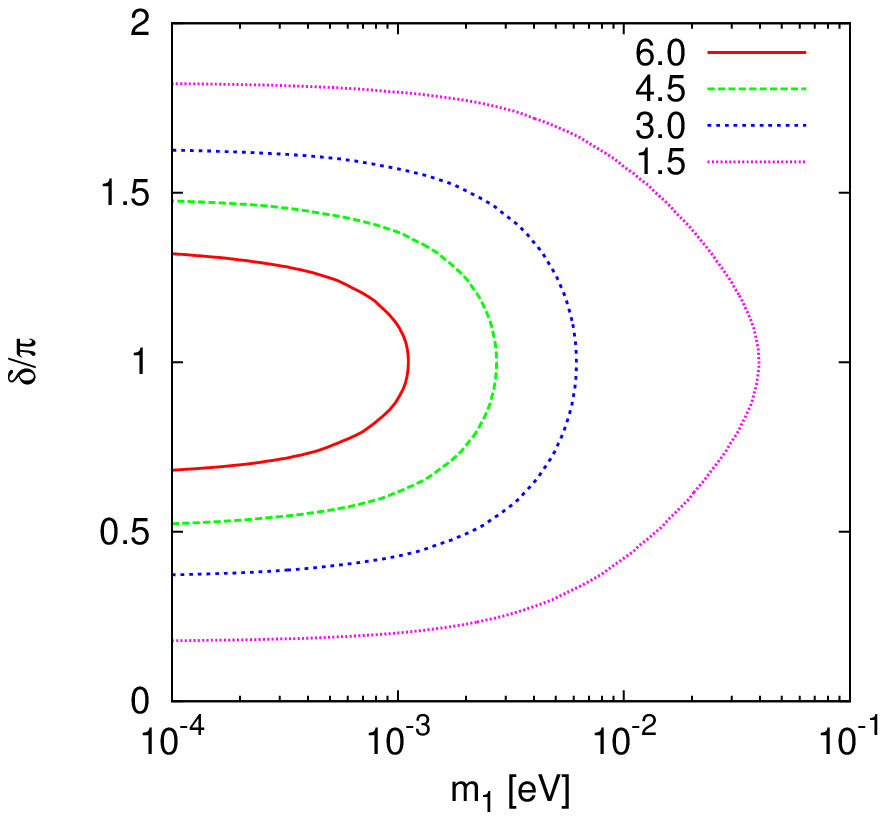}
\caption{$\text{Br}(\ell_\alpha \to \ell_\beta \gamma) /
  \text{Br}(\ell_\alpha \to \ell_\beta \gamma)_{\delta = 0}$ contours
  in the $m_{\nu_1} - \delta$ plane. To the left for $\ell_\alpha =
  \mu$ and $\ell_\beta = e$, to the right for $\ell_\alpha = \tau$ and
  $\ell_\beta = e$. Normal hierarchy for the light neutrinos, a
  degenerate right-handed neutrino spectrum and specific (but generic)
  values for the free parameters have been assumed, see text for more
  details.}
\label{fig:deltaRatios}
\end{figure}

As starting point, we discuss how a non-zero Dirac phase can change
the prediction for $\text{Br}(\ell_\alpha \to \ell_\beta \gamma)$. In
order to do that, we consider the ratio $\text{Br}(\ell_\alpha \to
\ell_\beta \gamma) / \text{Br}(\ell_\alpha \to \ell_\beta
\gamma)_{\delta = 0}$, where $\text{Br}(\ell_\alpha \to \ell_\beta
\gamma)_{\delta = 0}$ is the value of the branching ratio for $\delta
= 0$. This is explicitly shown in Fig. \ref{fig:deltaRatios}, where
contours of these ratios are drawn in the $m_{\nu_1} - \delta$
plane. In this figure we chose normal hierarchy for the light
neutrinos, a degenerate right-handed neutrino spectrum and a real $R$
matrix. Although these results were obtained for specific values of
the remaining parameters, we emphasize that the $\text{Br}(\ell_\alpha
\to \ell_\beta \gamma) / \text{Br}(\ell_\alpha \to \ell_\beta
\gamma)_{\delta = 0}$ does not depend on them when the right-handed
neutrinos are degenerate and $R$ is a real matrix, see
Eqs.~\eqref{LLG-yy} and \eqref{ydagy}.

The largest variations are found for $\text{Br}(\mu \to e \gamma)$ and
$\text{Br}(\tau \to e \gamma)$, most directly affected by
$\delta$. For the former, we find that the branching ratio can be
reduced by almost an order of magnitude, depending on the value of
$\delta$. In the latter case, the branching ratio can be increased by
a factor of $4$ just by switching on the Dirac phase. Moreover, in
both cases we find that $m_{\nu_1}$ is also determinant. We do not
show our results for the remaining case, $\ell_\alpha = \tau$ and
$\ell_\beta = \mu$, since we found very little dependence on the Dirac
phase.

\begin{figure}[t]
\centering
\includegraphics[width=0.49\linewidth]{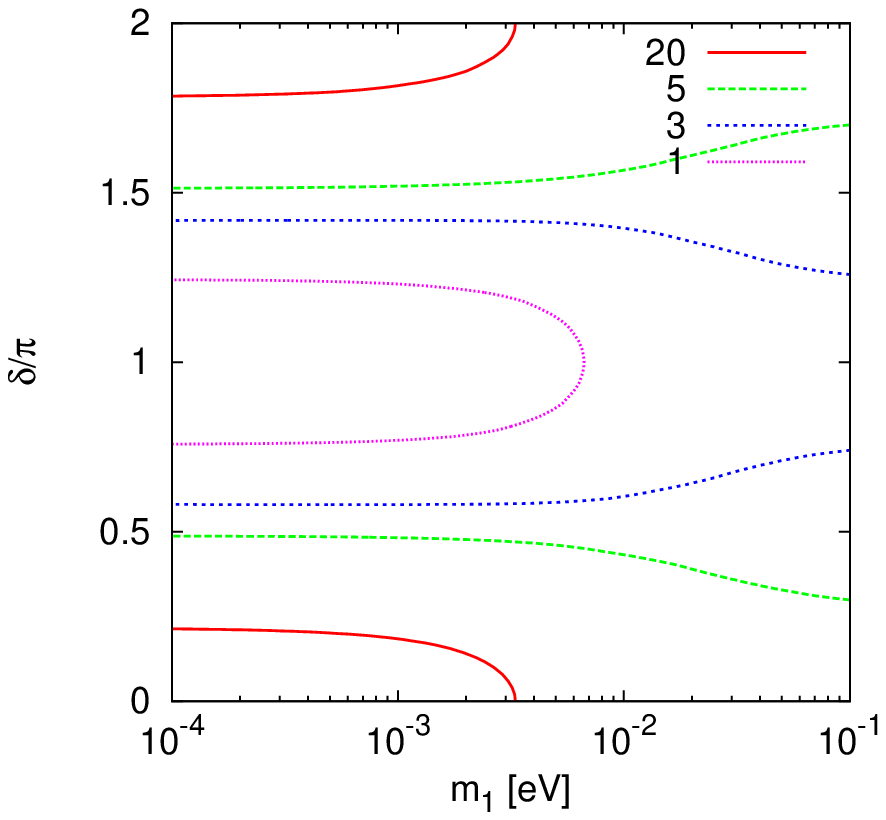}
\includegraphics[width=0.49\linewidth]{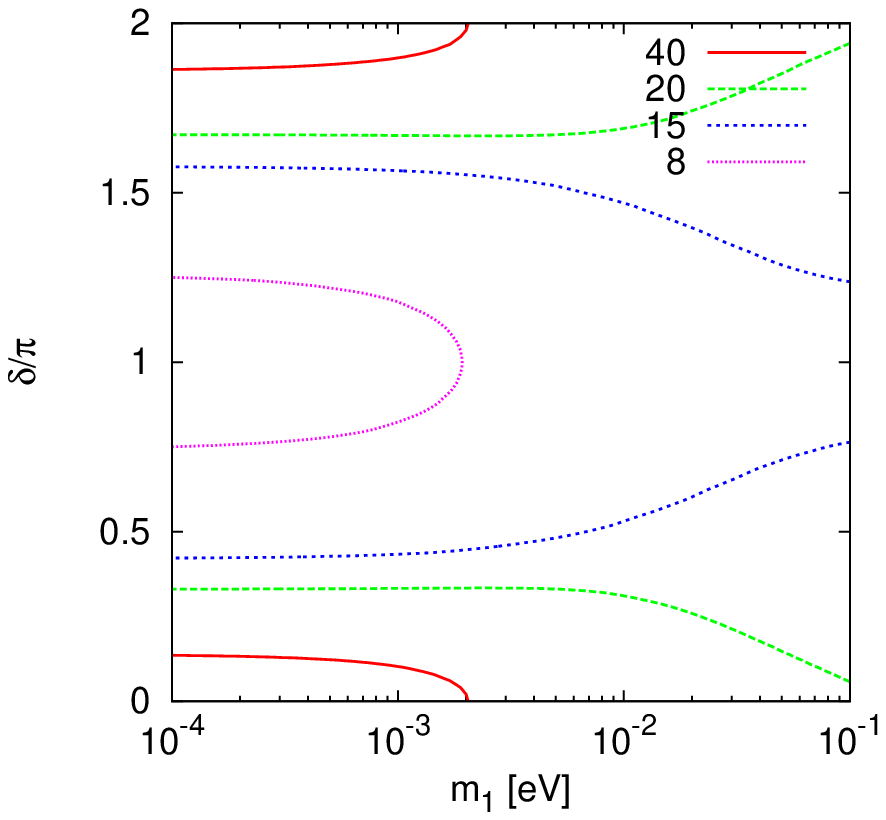}
\caption{$\text{Br}(\ell_\alpha \to \ell_\beta \gamma) /
  \text{Br}(\ell_{\alpha'} \to \ell_{\beta'} \gamma)$ contours in the
  $m_{\nu_1} - \delta$ plane. To the left for $\ell_\alpha = \mu$,
  $\ell_{\alpha'} = \tau$ and $\ell_\beta = \ell_{\beta'} = e$, to the
  right for $\ell_\alpha = \ell_{\alpha'} = \tau$, $\ell_\beta = \mu$
  and $\ell_{\beta'} = e$. Normal hierarchy for the light neutrinos, a
  degenerate right-handed neutrino spectrum and specific (but generic)
  values for the free parameters have been assumed, see text for more
  details.}
\label{fig:LLGRatios}
\end{figure}

These results tell us that the LFV rates are highly sensitive to the
low-energy neutrino parameters. The question then arises as to whether
one can get information about them by measuring LFV observables. In
case of $\text{Br}(\ell_\alpha \to \ell_\beta \gamma)$, we have
already seen that, for the specific scenario of degenerate
right-handed neutrinos and a real $R$ matrix, the flavor dependence of
the amplitude will be determined just by low energy parameters:
neutrino masses, mixing angles and CP violating phases. Therefore, by
taking ratios of branching ratios (what we call \textit{flavor
  ratios}), the dependence on the high-energy parameters cancels out
and we are left with functions of $m_{\nu_1}$ and $\delta$. More
precisely, we can make use of Eqs.~\eqref{LLG-yy} and \eqref{ydagy} to
write
\begin{equation}
\frac{\text{Br}(\ell_\alpha \to \ell_\beta
 \gamma)}{\text{Br}(\ell_{\alpha'} \to \ell_{\beta'} \gamma)} =
\frac{\left| \left( U_{\mathrm{PMNS}} \, \hat m_\nu \, U_{\mathrm{PMNS}}^\dagger
  \right)_{\beta \alpha}\right|^2}
{\left| \left( U_{\mathrm{PMNS}} \, \hat m_\nu \, U_{\mathrm{PMNS}}^\dagger
   \right)_{\beta' \alpha'} \right|^2} 
\frac{\text{Br}(\ell_\alpha\to \ell_\beta\nu_\alpha\overline{\nu_\beta})}
{\text{Br}(\ell_{\alpha'}\to \ell_{\beta'}\nu_{\alpha'}\overline{\nu_{\beta'}})}
\, . \label{eq:ratioLLG} 
\end{equation}
Note that there is no sum over $\alpha$, $\alpha'$, $\beta$ and
$\beta'$ in the previous expression.

Our results for these ratios are presented in
Fig. \ref{fig:LLGRatios}. We show $\text{Br}(\mu \to e \gamma) /
\text{Br}(\tau \to e \gamma)$ (to the left) and $\text{Br}(\tau \to
\mu \gamma) / \text{Br}(\tau \to e \gamma)$ (to the right) contours in
the $m_{\nu_1} - \delta$ plane. Under the assumptions of a degenerate
right-handed neutrino spectrum and vanishing phases in the $R$ matrix,
this figure would allow one to set important constraints on $m_{\nu_1}$
and $\delta$ in case two branching ratios were measured. Furthermore,
we see again the important dependence on these two low-energy
parameters, since the ratios can change by more than one order of
magnitude.

The same will be true for $\text{Br}(\ell_\alpha \to 3 \, \ell_\beta)$
when one of the two pieces, $D_1$ or $D_2$, dominates. A particularly
interesting scenario arises when the term containing the loop function
$D_2 \subset B$ gives the dominant contribution. As we have found
numerically, this assumption is typically valid for $\xi \gtrsim 10$
or for large $m_{\nu_1}$. In this case, the special dependence on the
Yukawa matrices, $\left( y^T y \right)_{\beta \beta} \left( y^T y
\right)_{\beta \alpha}$, implies that the $R$ matrix drops from the
flavor ratios even when it contains complex entries, since $R^T R =
1$.

\begin{figure}[t]
\centering
\includegraphics[width=0.49\linewidth]{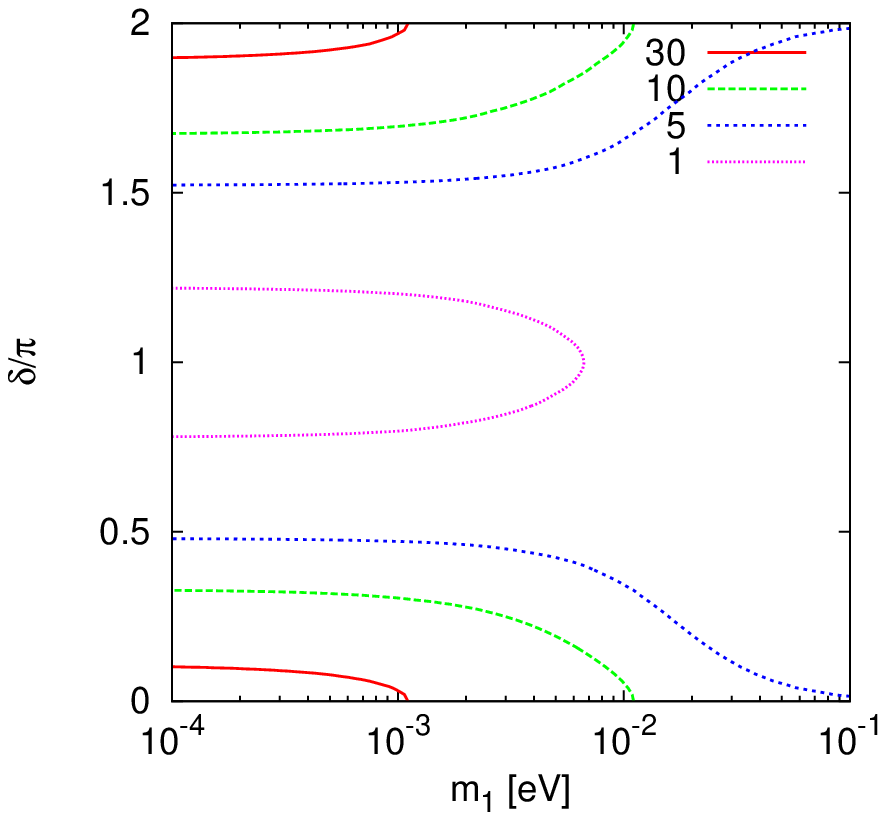}
\includegraphics[width=0.49\linewidth]{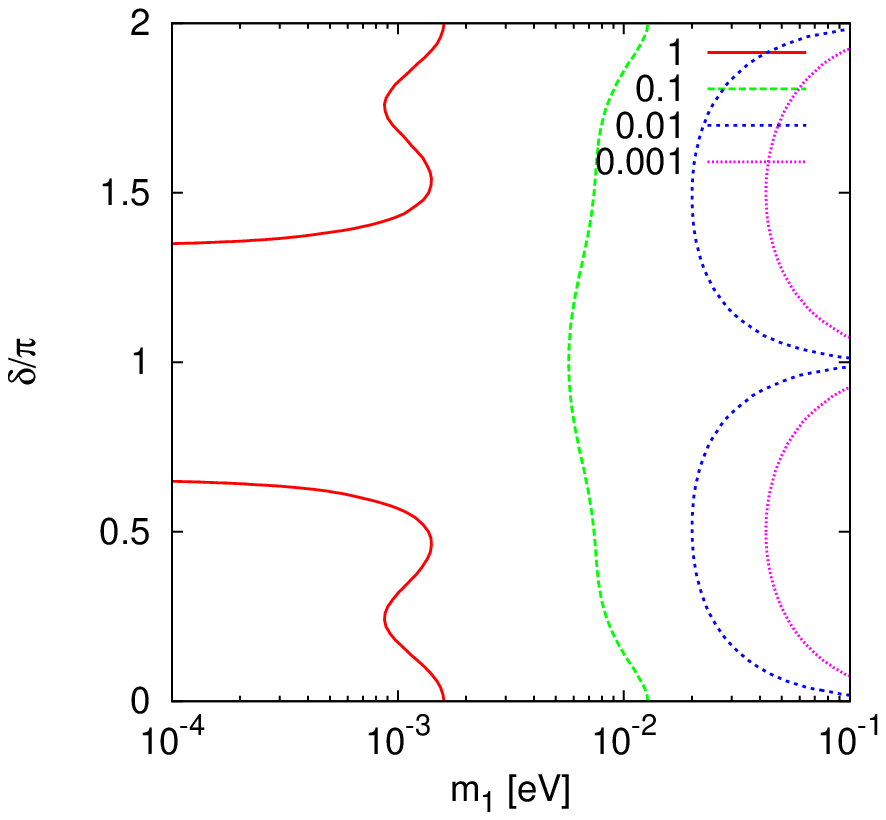}
\caption{$\text{Br}(\ell_\alpha \to 3 \, \ell_\beta) /
  \text{Br}(\ell_{\alpha'} \to 3 \, \ell_{\beta'})$ contours in the
  $m_{\nu_1} - \delta$ plane. To the left for $\ell_\alpha = \mu$,
  $\ell_{\alpha'} = \tau$ and $\ell_\beta = \ell_{\beta'} = e$, to the
  right for $\ell_\alpha = \ell_{\alpha'} = \tau$, $\ell_\beta = \mu$
  and $\ell_{\beta'} = e$. Normal hierarchy for the light neutrinos
  and a degenerate right-handed neutrino spectrum haven assumed, see
  text for more details.}
\label{fig:L3LRatios}
\end{figure}

We have investigated this scenario and obtained the results in
Fig.~\ref{fig:L3LRatios}. We concentrate on $\text{Br}(\mu \to 3 e) /
\text{Br}(\tau \to 3 e)$ (on the left) and $\text{Br}(\tau \to 3 \mu)
/ \text{Br}(\tau \to 3 e)$ (on the right). In the derivation of these
plots, we neglected the contribution from the $D_1$ term. Moreover, we
assumed normal hierarchy for the light neutrinos and a degenerate
right-handed neutrino spectrum. It is clear that, again, the
parameters $\delta$ and $m_{\nu_1}$ may have a very strong impact on
the 3-body branching ratios. On the left-hand side of the figure we
see that (for this parameter configuration) $\text{Br}(\mu \to 3 e)$
is typically larger than $\text{Br}(\tau \to 3 e)$. The ratio between
these two observables is only close to $1$ for $\delta = \pi$, whereas
in the rest of the $m_{\nu_1} - \delta$ plane one has $\text{Br}(\mu
\to 3 e) \gg \text{Br}(\tau \to 3 e)$. On the other hand, the
right-hand side of the figure shows that the ratio $\text{Br}(\tau \to
3 \mu) / \text{Br}(\tau \to 3 e)$ is mostly determined by $m_{\nu_1}$,
with $\delta$ playing a secondary role. As for the previous case, the
ratio could be close to $1$ (for low $m_{\nu_1}$) or much larger (for high
values of the lightest neutrino mass).

Our study reveals that LFV observables in the scotogenic model are
highly sensitive to low-energy parameters such as the Dirac phase or
the lightest neutrino mass. However, it also reveals a large
degeneracy, this is, the LFV rates are not correlated with a single
parameter. Furthermore, our results regarding flavor ratios have been
obtained for a special case: degenerate right-handed neutrinos and
real $R$ matrix. In a more general scenario one expects departures
from the values of the flavor ratios obtained here. In conclusion, it
is not possible to determine the value of a single parameter by
measuring a flavor ratio. Only the combination of measurements of the
low-energy parameters with the discovery of one (or several) LFV
processes can really put the flavor structure of the scotogenic model
under experimental test.

\subsection{$\mu-e$ conversion in nuclei}
\label{subsec:mueconv}

\begin{figure}[t]
\centering \includegraphics[width=0.6\linewidth]{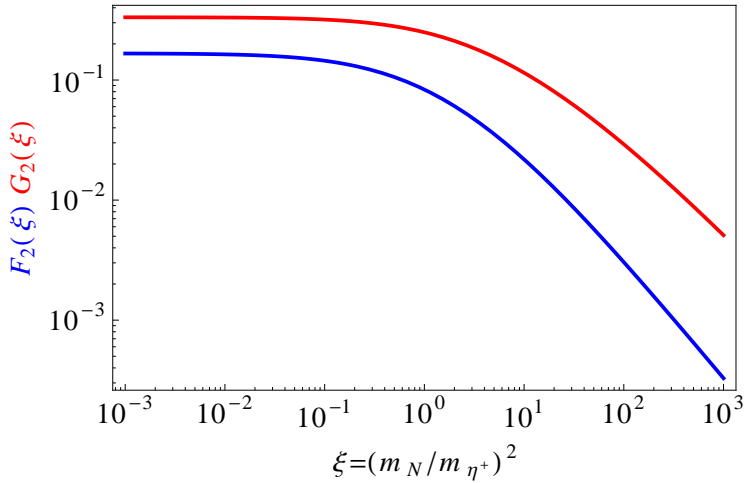}
\caption{The loop functions $F_2(\xi)$ and $G_2(\xi)$ as a function of
  $\xi = (m_N/m_{\eta^+})^2$. For the definitions see appendix
  \ref{sec:appendix1}.}
\label{fig:F2G2xi}
\end{figure}

\begin{figure}[t]
\centering
\includegraphics[width=0.49\linewidth]{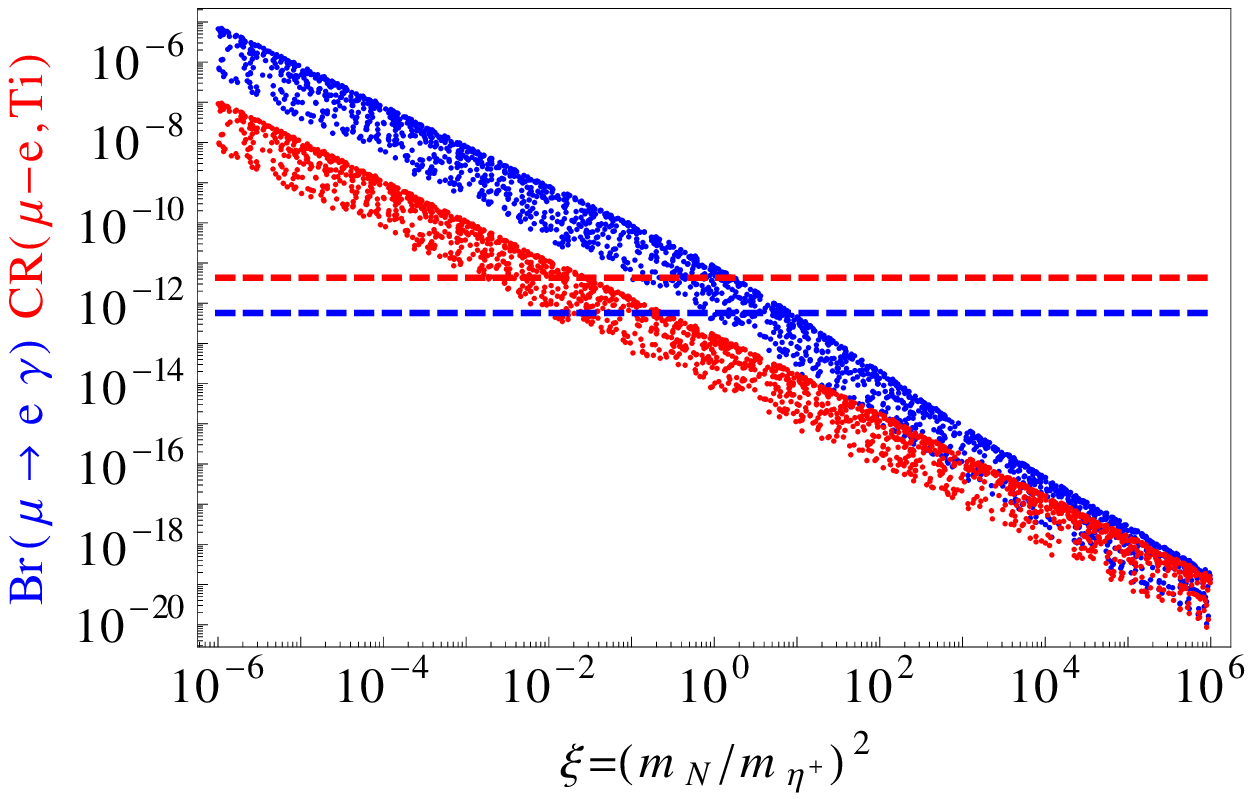}
\includegraphics[width=0.49\linewidth]{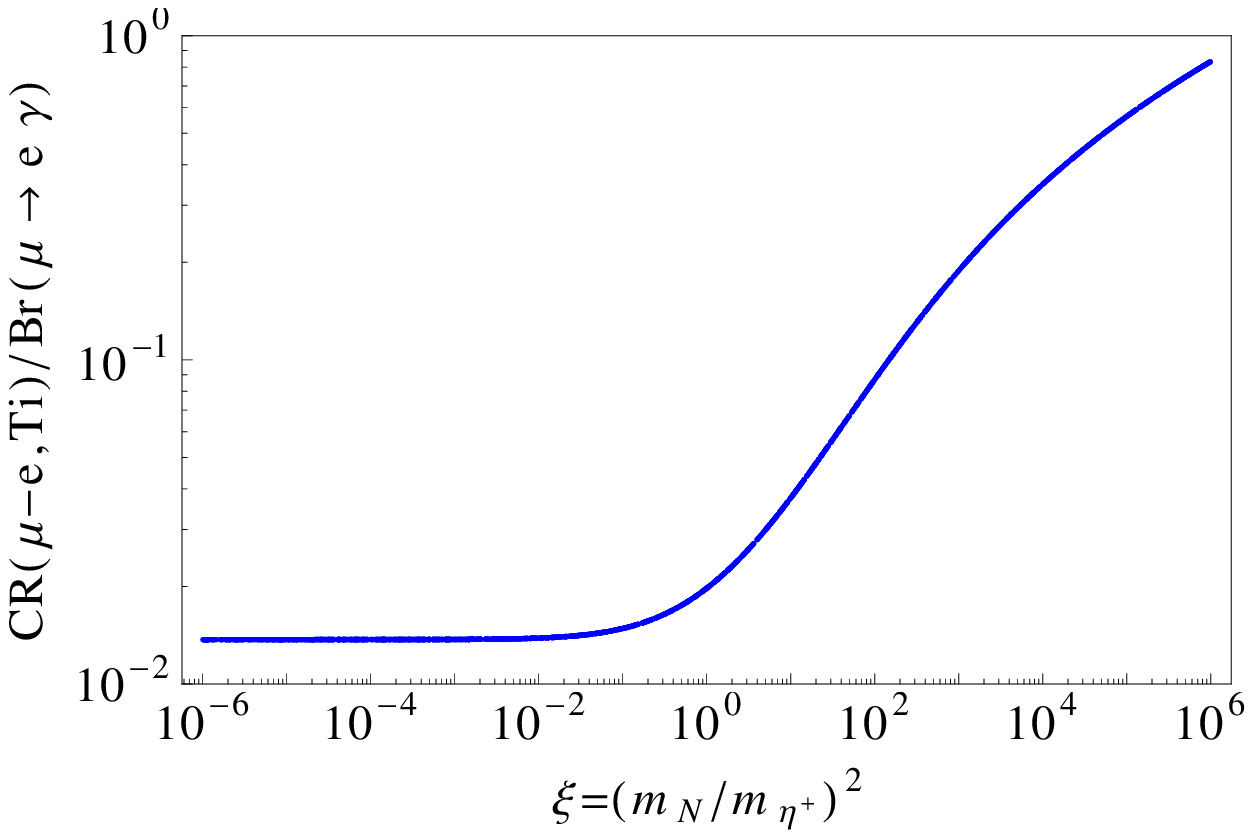}
\caption{$\text{Br}(\mu \to e \gamma)$ and ${\rm CR} (\mu- e, {\rm
    Ti})$ (to the left) and the ratio ${\rm CR} (\mu- e, {\rm
    Ti})/\text{Br}(\mu \to e \gamma)$ (to the right) as a function of
  $\xi = (m_N/m_{\eta^+})^2$. Normal hierarchy for the light neutrinos and
  a degenerate right-handed neutrino spectrum have been assumed, see
  text for details. The horizontal dashed lines show the current upper
  bounds.}
\label{fig:mueconv-xi}
\end{figure}

So far we have discussed our results on $\ell_\alpha \to \ell_\beta
\gamma$ and $\ell_\alpha \to 3 \, \ell_\beta$. Now we move on to
discuss $\mu-e$ conversion in nuclei.
In this model, we have found that $Z$-penguins give a very little
contribution to $\mu-e$ conversion in nuclei compared to that of the
$\gamma$-penguins. In this situation one could naively expect dipole
operators to dominate the conversion rate. When this is the case, one
expects a simple relation~\cite{Albrecht:2013iba}
\begin{equation}
\label{eq:mueconv-dipole}
\frac{{\rm CR} (\mu- e, {\rm Nucleus})}{\text{Br}(\mu \to e \gamma)}
\approx \frac{f(Z,N)}{428}
\end{equation}
where $f(Z,N)$ is a function that depends on the nucleus and ranges
from $1.1$ to $2.2$ for the nuclei of interest. However, in addition
to the dipole contribution given by $A_D$, $\gamma$-penguins also have
the non-dipole contribution given by $A_{ND}$. In fact,
Eqs. \eqref{eq:A2R} and \eqref{eq:A1L} tell us that, for degenerate
right-handed neutrinos, one has $A_D = 3 \, F_2(\xi)/G_2(\xi)
A_{ND}$. Therefore, the relative weight of these two different photon
contributions depends on the loop functions $F_2(\xi)$ and
$G_2(\xi)$. These are shown in Fig. \ref{fig:F2G2xi}, where one can
see that $G_2(\xi) > F_2(\xi)$. For $\xi \ll 1$ the difference between
$G_2(\xi)$ and $F_2(\xi)$ is small and both contributions have similar
weights. However, for $\xi \gg 1$ ($m_N \gg m_{\eta^+}$) one has
$G_2(\xi) \gg F_2(\xi)$ and $A_{ND}$ becomes the most relevant
contribution\footnote{Notice that we do not find the same behavior in
  $\text{Br}(\ell_\alpha \to 3 \, \ell_\beta)$ due to the additional
  (large) logarithmic factor that multiplies $|A_D|^2$ in the
  branching ratio formula, see Eq. \eqref{eq:l3lBR}. Moreover, even if
  the photonic non-dipole terms can be slightly larger than the dipole
  ones, box diagrams give even larger contributions in the same region
  of parameter space.}. This is illustrated in
Fig. \ref{fig:mueconv-xi}, where we show our results for
$\text{Br}(\mu \to e \gamma)$ and ${\rm CR} (\mu- e, {\rm Ti})$, as
well as their ratio. The same parameter configuration as in
Fig. \ref{fig:BRs-xi} has been selected: fixed values $m_{\eta^+} = 1$
TeV and $m_{\nu_1} = 10^{-3}$ eV, random real $R$ matrix and Dirac
phase. These numerical results were obtained for NH, although very
similar results are found for IH. We focused on $\mu-e$ conversion in
titanium, although the same behavior is found for other nuclei.

We find that for large values of $\xi$, the $\mu-e$ conversion rate in
titanium gets enhanced by photonic non-dipole contributions. This is a
positive result, given the great experimental perspectives for $\mu-e$
conversion in nuclei in the near future.

\section{Summary and conclusions}
\label{sec:conclusions}

The scotogenic model is a popular extension of the standard model that
accounts for neutrino masses and dark matter. As for most neutrino
mass models, lepton flavor violation is one of the most attractive
phenomenological issues, as it may reveal the underlying mechanism
that leads to neutrino masses and mixings. In this work we have
studied the predictions obtained in the scotogenic model for the LFV
processes with the best experimental perspectives in the near future:
$\ell_\alpha \to \ell_\beta \gamma$, $\ell_\alpha \to 3 \, \ell_\beta$
and $\mu-e$ conversion in nuclei. Full analytical expressions have
been derived, going beyond the usual dipole dominance
approximation. Our computation includes, besides the dipole photon
penguin contribution, non-dipole photon contributions, $Z$-penguins as
well as box diagrams.

The full consideration of all contributions to LFV processes leads to
a very interesting picture. Given the rich LFV phenomenology in the
scotogenic model, we are sure that more complete studies can be
performed. Here we have explored some of the phenomenological
consequences of our analytical results. This may serve as a summary of
our main conclusions:

\begin{itemize}

\item Box diagrams dominate the LFV amplitudes in some parts of
  parameter space. This scenario leads to a deviation from the naive
  expectations obtained from the dipole dominance assumption and makes
  $\ell_\alpha \to 3 \, \ell_\beta$ more constraining than $\ell_\alpha
  \to \ell_\beta \gamma$.

\item The mass hierarchy between the right-handed neutrinos and the
  inert doublet scalars is of fundamental relevance for LFV
  observables. We have found that parameter points with large Yukawa
  couplings and $m_N \gg m_{\eta^+}$ or $m_N \ll m_{\eta^+}$ typically
  have enhanced box diagrams, thus leading to $\text{Br}(\ell_\alpha
  \to 3 \, \ell_\beta) > \text{Br}(\ell_\alpha \to \ell_\beta
  \gamma)$. This is caused by the particular behavior of the loop
  functions.

\item In the scotogenic model, there are two dark matter candidates:
  the lightest right-handed neutrino $N_1$ and the lightest neutral
  $\eta$ scalar ($\eta_R$ or $\eta_I$)~\cite{Ma:2006km}. When $\xi>1$,
  the lightest neutral $\eta$ constitutes the dark matter of the
  universe. Otherwise, $N_1$ is the dark matter
  particle~\cite{Kubo:2006yx, Sierra:2008wj, Suematsu:2009ww,
    Schmidt:2012yg}. In case of $N_1$ DM ($\xi<1$), the only possible
  annihilation channel is $N_1N_1\to\ell_\alpha\bar{\ell}_\beta$, via
  the Yukawa interaction. For this reason, Yukawa couplings of
  $\mathcal{O}(1)$ are required in order to obtain the observed dark
  matter relic density $\Omega h^2\approx0.12$~\cite{Ade:2013zuv}, and
  this may lead to incompatibility with the LFV bounds. It is thus
  clear that the dark matter phenomenology of $N_1$ and LFV are
  closely related. We have explicitly constructed parameter points
  where all the requirements for right-handed neutrino dark matter are
  met: $m_N < m_\eta$, large Yukawa couplings and $m_N$ in the
  appropriate range, as found in dedicated
  studies~\cite{Schmidt:2012yg}. Our investigation reveals that
  although most of these points lead to violation of the LFV bounds, a
  small fraction of them are perfectly compatible. These valid points
  involve some small tuning of the parameters and could only be found
  due to the generality of our scans (not limited to any fixed
  structure of the Yukawa couplings). These results can be seen as a
  positive indication in favor of the validity of right-handed
  neutrino dark matter, although detailed studies are required to get
  a definitive and robust conclusion. These are, however, beyond the
  scope of this paper. On the other hand, we would like to point out
  that in case the dark matter is provided by the scalar $\eta$, one
  can always obtain the correct relic density since, in addition to
  the Yukawa interactions, this particle has gauge and scalar
  interactions~\cite{Kashiwase:2012xd, Kashiwase:2013uy}, not
  correlated with LFV.

\item The LFV rates are highly sensitive to the low-energy parameters
  $m_{\nu_1}$ (the mass of the lightest neutrino) and $\delta$ (the
  Dirac phase). In particular, large $m_{\nu_1}$ typically enhances
  box diagrams.

\item In some specific scenarios (with degenerate right-handed
  neutrinos), the ratios of branching ratios depend only on
  $m_{\nu_1}$ and $\delta$. Under some assumptions, this may allow us
  to test the flavor structure of the model.

\item Interestingly, the rate for $\mu-e$ conversion in nuclei can
  also be enhanced beyond the dipole contribution in some regions of
  the parameter space. Our study reveals that non-dipole photon
  contributions become very relevant for $m_N \gg m_{\eta^+}$. This
  may lead to $\mu-e$ conversion rates in nuclei as large as the
  branching ratio for $\mu \to e \gamma$. These are good news given
  the promising experimental projects in $\mu-e$ conversion in nuclei.

\end{itemize}

We would like to stress that our (qualitative) conclusions are not
restricted to Ma's scotogenic model, but should apply to a much wider
class of radiative neutrino mass models. In particular, extended
versions of the scotogenic model (like the model proposed in
\cite{Farzan:2009ji}) should have, at least in some corners of
parameter space, a similar phenomenology.

The presence of TeV scale particles with sizable couplings to the SM
states also leads to interesting prospects at the LHC. Although the
direct production of the right-handed neutrinos is typically
suppressed due to their singlet nature, they will be produced in the
decays of the $\eta$ scalars when this is kinematically allowed. In
turn, the $\eta$ scalars may have non-negligible production
cross-sections provided they are light. This possibility, not related
to the lepton sector, has been studied in some detail. In this case
one expects multilepton final states with a significant amount of
missing energy~\cite{Gustafsson:2012aj}. Furthermore, the scotogenic
states may also modify the usual Higgs boson decays, with observable
implications at the LHC~\cite{Ho:2013hia,Arhrib:2013ela}.

To conclude, the anatomy of lepton flavor violation in the scotogenic
model has been fully determined and some interesting phenomenological
aspects have been explored. Some definite predictions have been made,
and these may be used to put the model under experimental test. The
connection between neutrino masses and lepton flavor violation is a
powerful test for this purpose. Hopefully, a positive signal in one
(or several) experiments in the next few years will provide valuable
hints on the mechanism behind neutrino masses.

\section*{Acknowledgements}

We would like to thank Thomas Schwetz for many fruitful
discussions. AV also thanks Nuria Rius and Juan Racker for their
comments on the manuscript and acknowledges partial support from the
ANR project CPV-LFV-LHC {NT09-508531}.  TT acknowledges support from
the European ITN project (FP7-PEOPLE-2011-ITN,
PITN-GA-2011-289442-INVISIBLES).

\appendix

\section{Loop functions}
\label{sec:appendix1}
We present in this appendix the loop functions that appear in the paper,
\begin{eqnarray}
F_2(x) &=& \frac{1-6x+3x^2+2x^3-6x^2 \log x}{6(1-x)^4}, \\
G_2(x) &=& \frac{2-9x+18x^2-11x^3+6x^3 \log x}{6(1-x)^4}, \\
D_1(x,y) &=& - \frac{1}{(1-x)(1-y)} - \frac{x^2 \log x}{(1-x)^2(x-y)} -
 \frac{y^2 \log y}{(1-y)^2(y-x)}, \\
D_2(x,y) &=& - \frac{1}{(1-x)(1-y)} - \frac{x \log x}{(1-x)^2(x-y)} -
 \frac{y \log y}{(1-y)^2(y-x)}.
\end{eqnarray}
These loop functions do not have any poles. In the limit $x,y\to1$ and
$y\to x$, the functions become 
\begin{eqnarray}
F_2(1)&=&\frac{1}{12},\quad
G_2(1)=\frac{1}{4},\quad
D_1(1,1)=-\frac{1}{3},\quad
D_2(1,1)=\frac{1}{6},
\end{eqnarray}
\begin{eqnarray}
&&D_1(x,x)=\frac{-1+x^2-2x\log{x}}{(1-x)^3},\\
&&D_1(x,1)=D_1(1,x)=\frac{-1+4x-3x^2+2x^2 \log{x}}{2(1-x)^3},\\
&&D_2(x,x)=\frac{-2+2x-(1+x)\log{x}}{(1-x)^3},\\
&&D_2(x,1)=D_2(1,x)=\frac{1-x^2+2x\log{x}}{2(1-x)^3}.
\end{eqnarray}

\section{Flavor structures}
\label{sec:appendix2}

Using the conventions in Eq. \eqref{eq:PMNS} and neglecting the
Majorana phases one finds

\begin{itemize}
\item $U_{\mathrm{PMNS}} \, \hat{m}_\nu \, U_{\mathrm{PMNS}}^{\dag}$
\end{itemize}

\vspace*{-1.0cm}

\begin{eqnarray}
\left( U_{\mathrm{PMNS}} \, \hat{m}_\nu \, U_{\mathrm{PMNS}}^{\dag} \right)_{11} &=& c_{13}^2 (c_{12}^2 m_1+m_2 s_{12}^2)+m_3 s_{13}^2, \\
\left( U_{\mathrm{PMNS}} \, \hat{m}_\nu \, U_{\mathrm{PMNS}}^{\dag} \right)_{22} &=& s_{23}^2 \left[ s_{13}^2 (c_{12}^2 m_1+m_2 s_{12}^2)+c_{13}^2 m_3 \right] +c_{23}^2 (c_{12}^2 m_2+m_1 s_{12}^2) \label{eq:dag22} \\
&& + 2 \, c_{12} c_{23} s_{12} s_{13} s_{23} \cos \delta \, (m_1-m_2), \nonumber \\
\left( U_{\mathrm{PMNS}} \, \hat{m}_\nu \, U_{\mathrm{PMNS}}^{\dag} \right)_{33} &=& c_{23}^2 \left[ s_{13}^2 (c_{12}^2 m_1+m_2 s_{12}^2)+c_{13}^2 m_3 \right] +s_{23}^2 (c_{12}^2 m_2+m_1 s_{12}^2) \\
&& + 2 \, c_{12} c_{23} s_{12} s_{13} s_{23} \cos \delta \, (m_2-m_1), \nonumber \\
\left( U_{\mathrm{PMNS}} \, \hat{m}_\nu \, U_{\mathrm{PMNS}}^{\dag} \right)_{21} &=& c_{12} c_{13} s_{12} c_{23} (m_2-m_1)+c_{13} s_{13} s_{23} e^{-i \delta} \left[ m_3 - m_2 + c_{12}^2 (m_2-m_1) \right], \\
\left( U_{\mathrm{PMNS}} \, \hat{m}_\nu \, U_{\mathrm{PMNS}}^{\dag} \right)_{31} &=& c_{12} c_{13} s_{12} s_{23} (m_1-m_2)+c_{13} s_{13} c_{23} e^{-i \delta} \left[ m_3 - m_2 + c_{12}^2 (m_2-m_1) \right], \\
\left( U_{\mathrm{PMNS}} \, \hat{m}_\nu \, U_{\mathrm{PMNS}}^{\dag} \right)_{32} &=& c_{23} s_{23} \left[ (s_{12}^2 - c_{12}^2 s_{13}^2)(m_2 - m_1) + c_{13}^2 (m_3 - m_2) \right] \\
&& - c_{12} s_{12} s_{13} (c_{23}^2 e^{i \delta} - s_{23}^2 e^{-i \delta})(m_2 - m_1). \nonumber
\end{eqnarray}

\begin{itemize}
\item $U_{\mathrm{PMNS}}^* \, \hat{m}_\nu \, U_{\mathrm{PMNS}}^\dagger$
\end{itemize}

\vspace*{-1.0cm}

\begin{eqnarray}
\left( U_{\mathrm{PMNS}}^* \, \hat{m}_\nu \, U_{\mathrm{PMNS}}^\dagger \right)_{11} &=&  c_{13}^2 (c_{12}^2 m_1+m_2 s_{12}^2)+e^{-2 i \delta} m_3 s_{13}^2, \\
\left( U_{\mathrm{PMNS}}^* \, \hat{m}_\nu \, U_{\mathrm{PMNS}}^\dagger \right)_{22} &=&  s_{23}^2 \left[ e^{2 i \delta} s_{13}^2 (c_{12}^2 m_1+m_2 s_{12}^2)+c_{13}^2 m_3 \right] +c_{23}^2 (c_{12}^2 m_2+m_1 s_{12}^2) \\
&& + 2 \, c_{12} c_{23} s_{12} s_{13} s_{23} e^{i \delta} \, (m_1-m_2), \nonumber \\
\left( U_{\mathrm{PMNS}}^* \, \hat{m}_\nu \, U_{\mathrm{PMNS}}^\dagger \right)_{33} &=& c_{23}^2 \left[ e^{2 i \delta} s_{13}^2 (c_{12}^2 m_1+m_2 s_{12}^2)+c_{13}^2 m_3 \right] +s_{23}^2 (c_{12}^2 m_2+m_1 s_{12}^2) \\
&& + 2 \, c_{12} c_{23} s_{12} s_{13} s_{23} e^{i \delta} \, (m_2-m_1), \nonumber \\
\left( U_{\mathrm{PMNS}}^* \, \hat{m}_\nu \, U_{\mathrm{PMNS}}^\dagger \right)_{21} &=& c_{12} c_{13} s_{12} c_{23} (m_2-m_1)+c_{13} s_{13} s_{23} e^{i \delta} \left[ e^{-2 i \delta} m_3 - m_2 + c_{12}^2 (m_2-m_1) \right], \nonumber\\\\
\left( U_{\mathrm{PMNS}}^* \, \hat{m}_\nu \, U_{\mathrm{PMNS}}^\dagger \right)_{31} &=& c_{12} c_{13} s_{12} s_{23} (m_1-m_2)+c_{13} s_{13} c_{23} e^{i \delta} \left[ e^{-2 i \delta} m_3 - m_2 + c_{12}^2 (m_2-m_1) \right], \nonumber\\\\
\left( U_{\mathrm{PMNS}}^* \, \hat{m}_\nu \, U_{\mathrm{PMNS}}^\dagger \right)_{32} &=& c_{23} s_{23} \left[ (s_{12}^2 - e^{2 i \delta} c_{12}^2 s_{13}^2)(m_2 - m_1) + c_{13}^2 (m_3 - e^{2 i \delta} m_2) + (e^{2 i \delta} -1 ) m_2 \right], \nonumber\\
&& - c_{12} s_{12} s_{13} e^{i \delta} (c_{23}^2 - s_{23}^2)(m_2 - m_1). 
\end{eqnarray}

\end{document}